\documentclass{article}

\usepackage{arxiv}

\usepackage[utf8]{inputenc} % allow utf-8 input
\usepackage[T1]{fontenc}    % use 8-bit T1 fonts
\usepackage{hyperref}       % hyperlinks
\usepackage{url}            % simple URL typesetting
\usepackage{booktabs}       % professional-quality tables
\usepackage{amsfonts}       % blackboard math symbols
\usepackage{nicefrac}       % compact symbols for 1/2, etc.
\usepackage{microtype}      % microtypography
\usepackage{lipsum}
\usepackage{graphicx}
\usepackage{orcidlink}
\graphicspath{ {./images/} }

\usepackage{bm} 
\usepackage{amsmath}
\usepackage{subfig}
\newcommand{\shortcite}{\cite}
\newcommand{\degree}{^\circ}

\title{WildFireGS: Physics-Based Wildfire Simulation in Large-Scale Semantics-Enriched Gaussian Splatting Forest Scenes}

\author{
 Nienke Driessen \\
  Delft University of Technology\\
  \texttt{n.c.a.driessen@student.tudelft.nl} \\
   \And
 Joris Rijsdijk \\
  Delft University of Technology\\
  \texttt{j.a.rijsdijk@tudelft.nl} \\
  \And
 Sören Pirk \\
  Kiel University\\
  \texttt{soeren.pirk@gmail.com} \\
  \And
 Wojtek Palubicki \\
  Adam Mickiewicz University in Poznan\\
  \texttt{wojciech.palubicki@amu.edu.pl} \\
  \And
 Dominik L. Michels \\
  King Abdullah University of Science and Technology (KAUST)\\
  \texttt{dominik.michels@kaust.edu.sa} \\
  \And
 Michael Weinmann \\
  Delft University of Technology\\
  \texttt{m.weinmann@tudelft.nl} \\
}

\begin{document}
\maketitle
%\renewcommand{\thefootnote}{\fnsymbol{footnote}}
%\footnotetext[1]{Equal contribution}

\begin{abstract}
Climate-driven environmental change is driving an increase in both the frequency and severity of wildfire events, making accurate simulation and prediction critical for effective risk mitigation and landscape management.
While recent physics-based wildfire models achieve high realism by explicitly simulating combustion, heat transfer, and fuel dynamics, they remain largely restricted to synthetic environments with complete and idealized knowledge of forest structure, limiting their applicability to real-world environments captured via aerial imagery.
To provide a pathway toward real-world wildfire digital twins derived directly from observational data, we present WildFireGS, a physics-based wildfire simulation framework operating directly on large-scale, semantics-enriched 3D Gaussian Splatting forest reconstructions.
Our approach bridges learning-based scene reconstruction and environmental simulation by augmenting Gaussian primitives with semantics and material properties that encode vegetation type and fuel characteristics.
We introduce a particle-based combustion model that operates natively on Gaussian representations, simulating ignition, heat transfer, combustion, and flame propagation across complex forest structures.
This enables direct physics-based simulation of fire behavior on reconstructed real-world environments, without requiring conversion to explicit meshes or volumetric grids.
We further demonstrate the modularity of WildFireGS through a rain-driven cooling mechanism in terms of an energy-sink process to realistically model fire containment.
Evaluations on synthetic scenes and real aerial forest captures show physically consistent wildfire behavior, reproducing characteristic dynamics including propagation scaling with vegetation density, wind velocity, and terrain slope.
In addition, we validate our model through novel firebreak experiments and biomass loss estimation.
\end{abstract}

%\begin{teaserfigure}
\begin{figure*}
  \includegraphics[width=\textwidth]{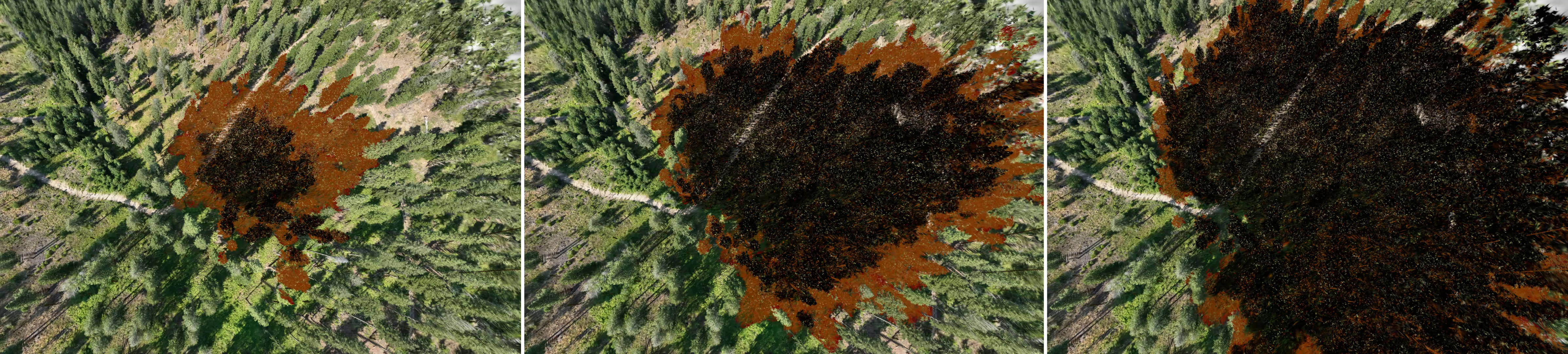}
  \caption{Snapshots from our simulation run on a real-world boreal-forest scene reconstructed from drone imagery captured by the Open Forest Observatory \cite{OpenForestObservatory-mission-001210-2025}. From left to right, the state of the burning scene is shown 10, 30, and 50 simulated minutes after ignition. The active fire front is represented by glowing orange Gaussians, while black Gaussians are fully burnt and retain no burnable mass; all remaining Gaussians keep their original appearance.}
  \label{fig:teaser}
\end{figure*}
%\end{teaserfigure}

\newcommand{\todo}[1]{{\textcolor{red}{TODO: #1}}}

% Math basics
\newcommand{\N}{\mathbb{N}}
\newcommand{\R}{\mathbb{R}}
\newcommand{\transpose}[1]{#1^\mathsf{T}}
\newcommand{\OLandau}{O}
\newcommand{\dInt}{\,\mathrm{d}}

% Work distribution
\newcommand{\gaussianCount}{G}
\newcommand{\gaussianPointCount}[1]{n_{#1}}
\newcommand{\gaussianThreadStart}[1]{t_{#1}}
\newcommand{\gaussianIndex}{g}
\newcommand{\gaussianRunningIndex}{i}
\newcommand{\threadIndex}{t}
\newcommand{\threadRunningIndex}{i}
\newcommand{\pointCountBound}{T}
\newcommand{\threadGaussianIndex}[1]{g_{#1}}

% Opacity correction
\newcommand{\cov}{\Sigma}
\newcommand{\invCov}{\cov^{-1}}
\newcommand{\mean}{\mu}
\newcommand{\opacity}{\alpha}
\newcommand{\pixelCoord}{q}
\newcommand{\gaussOpacity}{\alpha}
\newcommand{\poissonRate}{\lambda}
\newcommand{\pointCountRealization}{k}
\newcommand{\sampledDensity}{p}
\newcommand{\subpixelCenter}{q}
\newcommand{\subpixelArea}{A}

% Sampling strategies
\newcommand{\covSqrt}{L}
\newcommand{\unitCoord}{o}
\newcommand{\radius}{r}
\newcommand{\sampledCDF}[2]{F_{#1}(#2)}
\newcommand{\invSampledCDF}[2]{F_{#1}^{-1}(#2)}
\newcommand{\dilog}{\operatorname{Li}_2}
\newcommand{\invDilog}{\operatorname{Li}_2^{-1}}
\newcommand{\dilogIntegrationVariable}{s}
\newcommand{\randA}{u_0}
\newcommand{\randB}{u_1}
\newcommand{\sampledUnitCoord}{\unitCoord}
\newcommand{\sampledPixelCoord}{\pixelCoord}
\newcommand{\pointCountFactor}{K}

% appendix
\newcommand{\maxRadius}{\radius_{\max}}
\newcommand{\runningRadius}{\radius}

\section{Introduction\label{sec:introduction}}

Climate-driven environmental change is rapidly increasing the frequency, scale, and severity of wildfire activity worldwide~\cite{campen-2025,jones-2024}.
Rising global temperatures, prolonged droughts, declining fuel moisture, and increasingly extreme weather conditions contribute to extreme wildfire events across the globe.
In the 2023--2024 fire season alone, global wildfire activity burned approximately $3{.}9$ million square kilometers, with wildfire carbon emissions reaching $2{.}4$ Pg C, around $16\%$ above long-term averages~\cite{jones-2024}.
These escalating fire events pose growing threats to ecosystems, biodiversity, infrastructure, public health, and global carbon cycles, underscoring the urgent need for improved simulation and prediction tools capable of supporting prevention, suppression planning, ecological management, and long-term climate resilience.
Accurate wildfire simulation has therefore become increasingly critical for both scientific understanding and practical decision-making.
Predicting wildfire dynamics in complex real-world environments remains a fundamental challenge, due to the complexity arising from the coupled processes of combustion, heat transfer, fuel properties, fluid dynamics, vegetation structure (e.g., tree arrangement and density, canopy height), and environmental conditions that together determine how fires ignite, propagate, and evolve across natural landscapes.
Traditional approaches for wildfire modeling primarily relied on empirical relationships between fuel, wind, and terrain~\cite{rothermel1972mathematical,anderson1983predicting}, statistical modeling based on correlations between environment and fire outcomes~\cite{preisler2004probability,liu2013future,taylor2013wildfire,hernandez2015statistical} or cellular automata to model propagation through local interaction rules on discretized landscapes~\cite{bak1990forest-fire,clarke1994cellular,karafyllidis1997model,encinas2007simulation,ALEXANDRIDIS2008191,DBLP:journals/tomacs/TrunfioDRSG11,freire-2019,fire3030026}. 
Despite providing valuable insights into wildfire spread and risk, these models generally lack the detailed physical coupling of combustion, heat transfer, vegetation structure, and environmental dynamics required for realistic simulation of wildfire behavior in complex forest ecosystems.
In contrast to relying on empirical correlations or heuristic propagation rules, physics-based wildfire modeling \cite{Linn1997,linn2002firetec,mell2007grassland,coen-2013,vanella-2021,DBLP:journals/tog/HadrichBPPM21,DBLP:journals/tog/NielsenBSB22,DBLP:journals/tog/KokoszaWEMLMPP24} aims to capture the underlying physical processes governing wildfire spread through explicit representations of vegetation structure, combustion processes, heat transfer, and fluid–fire interactions.
Despite these advances, most physics-based wildfire simulators rely on fully specified or synthetic environments with known vegetation geometry, fuel distribution, and material properties, limiting direct applicability to real-world forests reconstructed from incomplete and uncertain aerial or satellite observations.
These limitations highlight the need for a unified framework that connects real-world forest reconstruction with physically grounded wildfire dynamics. Recent advances in learning-based scene representations \cite{DBLP:conf/eccv/NeRF,DBLP:journals/tog/GaussianSplatting} establish a powerful foundation for bridging this gap, with Gaussian Splatting providing an explicit, scalable, and semantically extensible representation capable of reconstructing large-scale forest environments from aerial imagery while preserving the structural detail necessary for downstream physical simulation.

In this paper, we present \textbf{\emph{WildFireGS}}, a physics-based wildfire simulation framework operating directly on large-scale, semantics-enriched Gaussian Splatting forest reconstructions derived from real-world aerial imagery.
Our method augments Gaussian primitives with semantic vegetation classes, material properties, and fuel characteristics, enabling physically grounded combustion simulation directly on reconstructed forest scenes.
We introduce a particle-based combustion model that natively simulates ignition, heat transfer, fuel consumption, and wildfire propagation within the Gaussian splatting reconstruction of the environment, without requiring conversion to meshes or volumetric grids.
Furthermore, we demonstrate the modularity of our \textbf{\emph{WildFireGS}} approach through a rain-driven cooling mechanism in terms of an energy-sink process to realistically model fire containment.
Experiments on synthetic environments and real aerial forest captures confirm that the model reproduces characteristic wildfire dynamics, including propagation behavior driven by vegetation density, wind, and terrain slope.
We further validate our approach through firebreak experiments and biomass loss estimation, providing additional evidence of its ability to reproduce realistic fire–fuel interaction behavior.

In summary, our main technical contributions are:
 \begin{enumerate}
     \item \textbf{Semantics-enriched Gaussian forest reconstructions.} We extend Feature-Splatting-based 3DGS reconstructions of aerial forest imagery with per-Gaussian semantic vegetation classes, material properties, and fuel characteristics, yielding a scene representation that is directly usable as input to a physics-based wildfire simulator.
     \item \textbf{Dedicated scene-processing pipeline.} We introduce a pre-processing stage that cleans the raw 3DGS reconstruction (outlier culling, scale and gravity alignment), regularises it via ground and canopy height maps, synthesises missing tree-stem geometry from aerial captures, and enforces physically motivated combustibility constraints.
     \item \textbf{Particle-based combustion model on Gaussians.} We propose a Lagrangian, particle-driven combustion model that operates natively on Gaussian primitives and couples ignition, radiative and convective heat transfer, temperature-dependent pyrolysis, and mass loss without requiring conversion to meshes or volumetric grids.
     \item \textbf{Modular environmental coupling via rain particles.} We show that the particle-based formulation generalises beyond fire by introducing rain particles as energy sinks, yielding a physically consistent extinguishment mechanism and demonstrating the modularity of the framework.
     \item \textbf{Evaluation on synthetic and real-world scenes.} We validate WildFireGS on a suite of controlled synthetic experiments (vegetation density, terrain slope, wind speed, simulation consistency) and on in-the-wild aerial captures, including novel firebreak and biomass-loss experiments beyond the standard wildfire-simulation evaluation protocol. 
 \end{enumerate}

\section{Related Work\label{sec:related_work}}

\paragraph{\textbf{
Wildfire modeling and simulation:}} 
Early approaches for wildfire modeling focused on describing fire behavior through empirical relationships between fuel, wind, and slope and remain foundational in operational fire prediction~\cite{rothermel1972mathematical,anderson1983predicting}.
Further approaches have been built on statistical modeling to model correlations between environment and fire outcomes~\cite{preisler2004probability,liu2013future,taylor2013wildfire,hernandez2015statistical,coen-2013} and cellular automata for spatial fire spread simulation to model propagation through local interaction rules on discretized landscapes~\cite{bak1990forest-fire,clarke1994cellular,karafyllidis1997model,encinas2007simulation,ALEXANDRIDIS2008191,DBLP:journals/tomacs/TrunfioDRSG11,freire-2019,fire3030026}.
Further extensions focused on integrating probabilistic transition rules and machine-learning-based components to improve adaptability and predictive performance in complex environments~\cite{zheng2017forest,f13121974,DBLP:conf/ascat/GhoshAC24}.
Even though these models provide valuable insights into wildfire risk and propagation patterns, they generally remain limited in their ability to represent the full spatial and physical complexity of fire behavior, lacking the detailed coupling of combustion, heat transfer, vegetation structure, and environmental dynamics required to accurately simulate wildfire evolution in realistic forest environments.
Building on these limitations, there has been a shift towards physics-based wildfire modeling \cite{Linn1997,linn2002firetec,mell2007grassland,coen-2013,vanella-2021,DBLP:journals/tog/HadrichBPPM21,DBLP:journals/tog/NielsenBSB22,DBLP:journals/tog/KokoszaWEMLMPP24}, where fire behavior is modeled through explicit representations of vegetation structure, combustion processes, heat transfer, and fluid–fire interactions~\cite{brown1974handbook,albini1976estimating}.
These approaches aim to capture the underlying physical processes governing wildfire spread rather than relying on empirical correlations or heuristic propagation rules.
However, most physics-based wildfire simulation systems remain fundamentally constrained by their reliance on fully specified or synthetically constructed environments, requiring complete knowledge of vegetation geometry, fuel distribution, and material properties, which is rarely available in real-world forest settings.
As a result, these methods are difficult to apply directly to environments reconstructed from aerial or satellite imagery, where occlusions, incomplete observations, and reconstruction uncertainty are inherent.

\paragraph{\textbf{Scene reconstruction and learning-based scene representations:}}
Accurate wildfire simulation in real environments depends on robust scene reconstruction.
While airborne LiDAR allows highly precise forestry mapping~\cite{fassnacht2024survey}, its reliance on specialized aircraft, sensors, and costly operational logistics constrains large-scale deployment, making scalable image-based reconstruction approaches increasingly attractive~\cite{csillik2019monitoring}.

Traditional 3D reconstruction pipelines use feature-based multi-view correspondence estimation and geometric reasoning, combining Structure-from-Motion (SfM) for camera pose and sparse scene estimation~\cite{ullman1979interpretation,schonberger2016sfm} with Multi-View Stereo (MVS) for dense reconstruction~\cite{snavely2006phototourism,furukawa2010,schonberger2016mvs}.
However, these methods often struggle with sparse viewpoints, occlusions, low-texture regions, and the structural complexity of natural vegetation, leading to incomplete or noisy geometry in natural environments.
Recent learning-based systems such as DUSt3R~\cite{wang2024dust3r} and VGGT~\cite{wang2025vggt} employ transformer architectures for feed-forward multi-view 3D scene reconstruction, substantially improving robustness and scalability.
Beyond such pre-trained geometry-focused priors, inverse rendering methods such as Neural Radiance Fields (NeRFs)~\cite{DBLP:conf/eccv/NeRF} represent scenes as continuous radiance and density fields optimized through differentiable volume rendering~\cite{kajiya1986rendering,max1995optical,novak2018monte}. While highly expressive for photorealistic view synthesis, their implicit parameterization lacks direct correspondence to local 3D structure, limiting interpretability and editability. In addition, dense ray sampling and repeated network evaluation impose substantial computational cost, restricting scalability for large outdoor environments and downstream physical simulation.

3D Gaussian Splatting (3DGS) replaces implicit neural fields with explicit anisotropic Gaussian primitives for efficient, spatially localized scene representation and fast rasterization-based rendering~\cite{DBLP:journals/tog/GaussianSplatting}, thereby significantly facilitating interpretability and editing operations. Semantic extensions further enrich these representations by augmenting the Gaussian primitives with material and object-level semantics~\cite{DBLP:conf/eccv/QiuYZW24}. Recent fire-focused Gaussian splatting systems demonstrate their potential for flame reconstruction and fire synthesis~\cite{shui2024FlameGS,nazarenus2025gaussiansonfire,jia2026Wildfire3D,shen2026fierygs}, but primarily target appearance and dynamic reconstruction rather than physically grounded wildfire propagation in semantically structured forest environments.

\paragraph{\textbf{Tree and forest representations:}} 
%
%3D plant representation methods
Plant representation methods can be categorized into global models that describe plants as aggregated entities (e.g., crowns or functional compartments), modular approaches that decompose plants into repeated structural units, and multi-scale representations that represent vegetation hierarchically to improve scalability for both rendering and simulation \cite{godin2000representing}.
Examples include plant-environment simulation frameworks with biophysical plug-ins~\cite{bailey2019helios} and digital twin approaches to model plant growth dynamics~\cite{DBLP:journals/cea/MitsanisHT24}. Conifer modeling further spans procedural and simplified geometric representations~\cite{DBLP:conf/graphicsinterface/MeyerN00,scrbbr:zhang2007}, physics-aware leaf combustion models~\cite{DBLP:journals/tog/PirkJHMP17}, and particle-based Lagrangian structures in conical volumes that align with point-based representations such as 3DGS~\cite{osti_1642826}.

Classical forest reconstruction relies on airborne LiDAR~\cite{fassnacht2024survey}, terrestrial laser scanning~\cite{trochta20173d}, photogrammetry~\cite{Nobuo_Kochi202221068,murtiyoso2024virtual}, and their fusion~\cite{qiu2023forest}. More recent approaches extend Gaussian-splatting-based methods to vegetation reconstruction. ForestSplat~\cite{shaheen2025ForestSplat} demonstrates large-scale forest monitoring from consumer-grade drone imagery with LiDAR-comparable canopy and structure estimation at lower cost, while further work enables fine-grained tree reconstruction for structural metrics such as branching and diameter estimation~\cite{rs17081473}. Further approaches include procedural ecosystem models to synthesize forests via growth and competition dynamics~\cite{DBLP:journals/tog/MakowskiHSMPP19} as well as coupled vegetation–soil–atmosphere feedback modeling for environmental realism~\cite{DBLP:journals/tog/PalubickiMGHMP22}.

\paragraph{\textbf{Physics-based combustion and environmental simulation:}} 
Fire and combustion simulation spans approaches with varying trade-offs between physical realism, computational cost, and scalability.
Early graphics methods rely on grid-based fluid solvers for reactive flow, turbulence, smoke, and heat transport~\cite{bridson2008fluid,stam1999stablefluids,fedkiw2001smoke,nguyen2002,hong2010geometry,yuen-2014}, enabling realistic flames via physically inspired~\cite{nguyen2002,pegoraro2006physically,zhou-2021}, flame-specific~\cite{nguyen2001}, or artistically controlled formulations \cite{lamorlette2002structural}, but remain computationally expensive due to dense volumetric discretization.
Particle-based methods offer a more scalable alternative by representing fire as stochastic particle systems with attributes such as velocity, lifespan, and combustion state~\cite{10.1145/800059.801167,wu-2011}, and were later extended via hybrid particle–grid models to improve turbulence and detail~\cite{horvath-2009}. However, these approaches primarily target visual appearance rather than physically grounded combustion or fuel consumption.
Combustion-specific models focus on explicit heat transfer and material degradation, ranging from early treatments of conduction, convection, and combustion products~\cite{1167889} to surface burning~\cite{chiba1994visual}, volumetric combustion~\cite{zhao2003voxels}, and propagating burn fronts on disconnected geometries~\cite{liu2012simulating}. Radiative heat transfer is incorporated to model long-range ignition and spread effects~\cite{deris2000radiation,hong2010geometry}, though often with simplified material heterogeneity and limited scalability. Further work addresses combustion and heat diffusion on articulated surface geometry~\cite{hong2010geometry}. While material-point and thermodynamic methods further improve material realism~\cite{stomakhin2014augmented,nielsen-2022}, they are typically not designed for large-scale wood combustion or wildfire propagation.
In forestry and wildfire research, models range from the simulation of heat transfer~\cite{encinas2007simulation}, charring~\cite{wichman1987charring,lizhong2022charring}, and pyrolysis of full plants~\cite{bohren1973} to species-dependent flammability~\cite{lawes2011}, canopy effects~\cite{schwilk2003flammability}, moisture-driven combustion \cite{masinda2021moisture}, coupled fire–atmosphere systems~\cite{coen2005simulation,sun2009wfs}, and conifer-specific combustion models~\cite{mendoza2019}.
Classical wildfire spread models include geometric fire-front formulations such as elliptical growth~\cite{richards1990elliptical}, mathematical and semi-empirical descriptions~\cite{pastor2003mathematical}, and biome-specific simulations~\cite{cheney1993,dupuy2000}. 
While these approaches provide valuable ecological insight, they often rely on simplified vegetation geometry or focus on prediction rather than interactive, scene-specific combustion.
Recent work increasingly integrates physically based combustion with structured vegetation models and large-scale wildfire dynamics~\cite{nielsen-2022,10.1145/3747855,DBLP:journals/tog/HadrichBPPM21,DBLP:journals/tog/KokoszaWEMLMPP24}, but remains largely limited to controlled or synthetic environments.

\vspace{4mm}
In contrast to prior simulators relying on synthetic scenes and Gaussian-based fire methods focused on reconstruction or visual synthesis, we introduce a unified framework for physics-based wildfire simulation on large-scale, semantics-enriched Gaussian forest reconstructions from real imagery. By combining physically-grounded combustion with modern scene representations, it enables a foundation for wildfire digital twins and ecological simulation.
\section{Method\label{sec:method}}

\begin{figure*}[t]
    \centering
    \includegraphics[width=\textwidth]{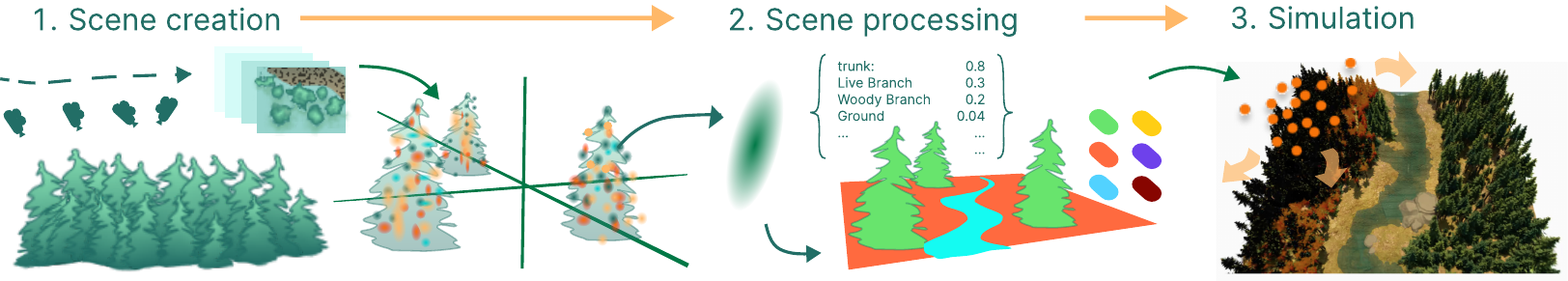}
    \caption{\textbf{\emph{WildFireGS Pipeline Overview}}. Starting from drone imagery, we reconstruct the scene via SfM and Feature Splatting, yielding a semantically enriched 3D Gaussian representation. Then, we pre-process the scene to prepare it for the simulation by filtering outliers, classifying Gaussian materials, and computing their mass. Finally, we run the physically based wildfire simulation, propagated through particles for both heat and rain transport.}
    \label{fig:pipeline}
\end{figure*}

We now present \textbf{\emph{WildFireGS}} as an effective approach for physics-based wildfire simulation in large-scale, semantics-enriched Gaussian Splatting forest reconstructions derived from real-world aerial imagery(Fig.~\ref{fig:pipeline}).
To enable physically grounded combustion simulation, we augment the Gaussian primitives with semantic vegetation classes, material properties, and fuel characteristics.
While Gaussians thereby define the state of the simulation, the lack of a connected geometric structure makes it non-trivial to propagate fire through the Gaussians via physical laws directly.
Instead, we opt for a Lagrangian combustion model, where physical laws are simulated on the Gaussian world state using particles as a medium.
While this allows us to natively simulate ignition, heat transfer, fuel consumption, and wildfire propagation within the Gaussian splatting reconstruction of the environment, it also offers the modularity for other purposes such as the introduction of cooling through rain particles.
In the following sections, we provide further details on the respective components of our approach.

\subsection{Scene Definition}\label{subsec:scene_definition}

Because combustion strongly depends on material properties~\cite{Khan2016}, we leverage Feature Splatting~\cite{DBLP:conf/eccv/QiuYZW24} to assign semantic material classes to individual Gaussians for simulation.

In addition to the standard 3DGS attributes (mean $\bm{\mu}$, scale $\mathbf{s}$, rotation $\mathbf{q}$, color $\mathbf{c}$, and opacity $\alpha$), each primitive is augmented with a state tuple $\{C, S, T, m\}$. The material class $C$ defines the flammability and overall physical behavior of the Gaussian, while the state $S \in \{S_{\mathrm{active}}, S_{\mathrm{burning}}, S_{\mathrm{burnt}}\}$ tracks its lifecycle. The temperature $T$ is modeled as a dynamic attribute dependent on the ambient scene temperature $T_{\mathrm{amb}}$ and the localized heat intensity of the fire. Furthermore, $m$ denotes the amount of burnable mass. Although $m$ could be modeled as the product of the perceived volume $V \propto s_xs_ys_z$ of a Gaussian and the density $\rho_C$ of its assigned material $C$, i.e., $m = V \cdot\rho_C$, we instead define it as a scene-wide average. This choice is necessitated by the fact that 3DGS typically produces a surface-aligned representation rather than an accurate volumetric model. Thus, naively computing the mass of a Gaussian as the product of $V$ and $\rho_C$ results in physically incorrect behavior.

Using Knud Thomsen’s formula with $p \approx 1.6075$ (max error $\pm 1.061\%$) \cite{michon-2005}, we compute the average surface area $\bar{A}$ for all $N$ Gaussians as:
\begin{equation}
\bar{A} = \frac{1}{N} \sum^N_{i=1} 4\pi \left( \frac{s_{x,i}^p s_{y,i}^p + s_{x,i}^p s_{z,i}^p + s_{y,i}^p s_{z,i}^p}{3} \right)^{\frac{1}{p}}\,.
\end{equation}

\paragraph{\textbf{Semantic Segmentation}}

Focusing on coniferous forests, we adopt the tree classes from the FOR-instance dataset \cite{DBLP:journals/corr/abs-2309-01279}, namely stem, woody branches, and live branches. These classes capture stems, bare branches, and foliage-bearing branches, providing a mesoscale representation that remains visually distinguishable at a distance while still being sufficiently distinctive to model the primary differences in combustion.

In addition, we consider a secondary list of classes (including rocks, water, and roads) that are nonflammable or offer different combustion characteristics compared to the tree classes defined above. We adapt this list per scene to accurately reflect the material contents while keeping the number of classes to a minimum.

To assign classes to each Gaussian, we perform a matching of each Gaussian’s learned feature vector from Feature Splatting against CLIP-encoded semantic text prompts~\cite{pmlr-v139-radford21a}, selecting the highest-confidence class above a similarity threshold (we use a confidence threshold of $5\%$), with uncertain primitives defaulting to non-flammable.
We found that including negative prompts in the classification process reduced the segmentation quality, particularly for the tree classes.

\subsection{Scene Processing}\label{subsec:scene_processing}
To prepare 3DGS scenes for fire simulation, we first cull Gaussians with extreme scales or low opacity values and apply a global transformation to align the scene with the gravitational axis. Subsequently, we perform the following steps.

\paragraph{\textbf{Spatial Mapping}}
We regularize the simulation space by computing 2D height maps (cell size $0.5\,$m) for both the ground and canopy layers. These maps are generated by rasterizing the means of class-specific Gaussians (ground Gaussians for the ground map, live branch Gaussians for the canopy map) as points and initializing cell heights through local averaging. For the ground map, we clip vertical coordinates at the $3^{\mathrm{rd}}$ and $97^{\mathrm{th}}$ percentiles to remove outliers. In both cases, we fill holes using a $k$-d tree nearest neighbor search and apply a low-pass Gaussian filter $G_{\sigma}$ for smoothing.

\paragraph{\textbf{Tree Stem Synthesis}}
Aerial captures often under-represent tree stem geometry relative to the canopy. To resolve this, we identify tree centers by detecting local maxima in the canopy height map (prominence $> 0.3\,$m). Candidates are filtered based on local Gaussian density and proximity to taller neighbors. For each validated tree, we insert a conical distribution of ``woody branch'' Gaussians extending from the ground to $80\%$ of the canopy height, following the structural model of Mendoza et al. \shortcite{osti_1642826}.

\paragraph{\textbf{Physical Constraints}}
We enforce combustibility overrides to prevent unrealistic fire propagation. Gaussians located below the ground height or within horizontal proximity to water-classified Gaussians are marked non-flammable. While our pipeline effectively handles scene geometry, persistent artifacts like sky or boundary Gaussians are pruned manually prior to simulation.

\subsection{Particle Dynamics}\label{subsec:particle_dynamics}
In our framework, we use particles to propagate physical processes through the scene and update the state of the Gaussians.
While Gaussians represent the combustible geometry, particles act as the dynamic medium for heat and mass transfer. Each particle $p$ is defined by a state tuple $\{\bm{\mu}_p, a_p, \mathbf{I}_p, l_p, r_p\}$, where $\bm{\mu}_p$ is the spatial position, $a_p$ is the current age, $\mathbf{I}_p$ is the intrinsic velocity, $l_p$ is the maximum lifetime, and $r_p$ is the radius of influence. To ensure natural variation in the simulation, $\mathbf{I}_p$ and $l_p$ are sampled from normal distributions during instantiation.

At each simulation timestep of size $\Delta t$\,, the particle age is advanced as $a_{p, t+1} = a_{p, t} + \Delta t$\,, and the particle is culled once $a_p \ge l_p$\,. Notably, our framework enforces a decoupled physics model where particles interact exclusively with Gaussians.
In particular, each Gaussian tracks a single change in energy $\Delta Q_{g,t}$ per frame from all particle interactions whose influence radius covers it.

\paragraph{\textbf{Fire Particles}}
We use fire particles that propagate thermal energy by augmenting standard particle attributes with temperature $T_p$ and temperature-dependent updraft velocity $\mathbf{v}_{\mathrm{up}}$ with $\|\mathbf{v}_{\mathrm{up}}\| \propto T_p$.
We instantiate these particles probabilistically, i.e., each Gaussian in the $S_{\mathrm{burning}}$ state has a defined probability $p_{\mathrm{spawn}}$ of releasing a particle per timestep.
The fire particles model the net heat flux $q_{\mathrm{net}} = q_{r} + q_{c}$ in $\mathrm{W}/\mathrm{m}^2$\,. Radiative heat flux $q_{r}$ is exchanged via electromagnetic radiation, modeled using the Stefan--Boltzmann law:
\begin{equation}
    q_{r} = \epsilon \sigma (T^4_{\mathrm{fire}} - T^4_{\mathrm{surface}})\,,
\end{equation}
where $T_{\mathrm{fire}}$ is the temperature of the emitting fire particle, $T_{\mathrm{surface}}$ the temperature of the receiving Gaussian, $\epsilon$ is the emissivity, and $\sigma$ the Stefan--Boltzmann constant. Convection $q_{c}$ is the heat flux within fluids, modeled as
\begin{equation}
    q_{c} = h (T_{\mathrm{fire}} - T_{\mathrm{surface}})\,,
\end{equation}
where $h$ is the convective heat transfer coefficient.
Each fire particle contributes a change of energy $\Delta Q = q_{\mathrm{net}} \bar{A}$ to all Gaussians within its influence radius.
The flux itself is modeled as the trajectory of a fire particle:
\begin{equation}
\bm{\mu}_{p, t+1} = \bm{\mu}_{p, t} + (\mathbf{v}_{\mathrm{wind}} + \mathbf{v}_{\mathrm{up}} + \mathbf{I}_p) \Delta t\,.
\end{equation}
The magnitude of $\mathbf{I}_p$ for a particle is positively correlated with its temperature. The scene-defined wind velocity $\mathbf{V}_{\mathrm{wind}}$ affects the horizontal drift of a particle as $\mathbf{v}_{\mathrm{wind}} = \eta \cdot \mathbf{V}_{\mathrm{wind}}$, with a dimensionless wind drift coefficient $\eta=0.1$.

To maintain a stable computational budget, we implement a dynamic population cap $N_{\mathrm{max}}$ that scales with the intensity of the fire. We interpolate $N_{\mathrm{max}}$ between a baseline (e.g., 500) and a peak (e.g., 5000) based on the current ratio of burning Gaussians to the total scene count. Available particle slots are filled by random sampling from the candidate pool generated in that timestep.

\paragraph{\textbf{Rain Particles}}
To demonstrate the modularity of the framework, we implement cooling particles to simulate rain. These particles are instantiated uniformly above the bounding box of the scene and descend to simulate precipitation. Each cooling particle carries a mass $m_p$ and a predefined thermal absorption capacity $Q$\,. This capacity is modeled as the energy required to raise the water to its boiling point and to account for the latent heat of vaporization:
\begin{equation}
Q = m_p c_w (T_{\mathrm{boil}} - T_w) + m_p L_v,
\end{equation}
where $T_{\mathrm{boil}}$ is the boiling point of water, $T_w$ the current temperature of the rain particle, $c_w$ is the specific heat capacity of water, and $L_v$ is the latent heat of vaporization.
The trajectory of a rain particle is modeled by
\begin{equation}
\bm{\mu}_{p, t+1} = \bm{\mu}_{p, t} + (\mathbf{v}_{\mathrm{wind}} + \mathbf{I}_p) \Delta t,
\end{equation}
where $\mathbf{I}_p$ is the terminal downward velocity of a rain droplet, and $\mathbf{v}_{\mathrm{wind}}$ is computed in the same way as for fire particles.

We use a greedy depletion mechanism: as a raindrop acting as a cooling particle traverses the scene, it interacts with the closest burning Gaussian $g$\,. Similar to fire particles, the energy of a Gaussian is reduced by $\Delta Q_g$\,. This process is only modeled by convection:
\begin{equation}
\Delta Q_g = h_c (T_g - T_w) \bar{A} \Delta t\,,
\end{equation}
where $T_g$ is the temperature of the affected Gaussian, $T_w$ the temperature of the rain particle, and $h_c = 150 \,\mathrm{W}\,\mathrm{m}^{-2}\,\mathrm{K}^{-1}$ is the heat transfer coefficient of a water condenser.
The equivalent energy is subtracted from the capacity of a rain particle: $Q_{t+1} = Q_{t} - \Delta Q_g$\,. If $Q \le 0$\,, the particle is culled as it is assumed to have transitioned to steam. This mechanism allows for a robust extinguishment system.

To reduce computational load, we combine many physically accurately sized particles into larger particles. Specifically, we spawn $n_p \propto P_r$ particles per unit of area per frame, depending on the precipitation rate $P_r$\,.
We compute particle mass as
\begin{equation}
    m_p = \rho_w \frac{\Delta t\, A_s\, P_r}{3600\, n_p}\,,
\end{equation}
given the total area of the ground in the scene $A_s$, the precipitation rate $P_r$ (in $\mathrm{mm}/\mathrm{h}$, the factor $3600$ converting hours to seconds), and the density of water $\rho_w$.
Unlike the fire particles which could grow exponentially in numbers, we do not need to explicitly enforce a limit due to the uniform manner in which rain is instantiated.

\subsection{Simulation Execution}\label{subsec:simulation_execution}
The simulation progresses through an iterative loop of particle updates and Gaussian state evaluations. In each step, we first update the particle positions and resolve their interactions with nearby Gaussians via a voxel-based acceleration hash structure. This allows for efficient proximity queries, ensuring that heat or cooling effects are only calculated for relevant Gaussian-particle pairs.

We use a fixed timestep of $\Delta t = 5\,\mathrm{s}$\,, which provides an optimal balance for capturing mesoscale simulation effects. This interval is sufficiently small to prevent particles from bypassing critical spatial interactions, yet large enough to maintain interactive performance across the entire simulation domain.
The particles accumulate an energy update per Gaussian $\Delta Q_{g,t}$, which is then used to update the energy of the Gaussian as $Q_{g, t+1} = Q_{g,t} + \Delta Q_{g,t}$. Given the mass of the Gaussian $m_g$ and the heat capacity $c_C$ of material class $C$, we compute the resulting change in temperature $\Delta T_g$ as
\begin{equation}
    \Delta T_g = \frac{Q_{g,t+1}} {m_g c_{C}}.
\end{equation}

Gaussians and water particles are initialized with $T_g = T_w = T_{\mathrm{amb}} = 15\,\degree \mathrm{C}\, $
and state $S_{\mathrm{active}}$. Once they have been heated beyond their material's ignition temperature, they are classified as $S_{\mathrm{burning}}$. Cooling below the extinguish temperature reverts them back to $S_{\mathrm{active}}$.
Upon exhausting all burnable mass, we set them to $S_{\mathrm{burnt}}$\,, after which they can no longer participate in the fire simulation.

\paragraph{\textbf{Gaussian Mass Loss}}
Mass loss represents the pyrolytic decomposition of the material as a result of a temperature-dependent reaction. 
We follow previous work~\cite{DBLP:journals/tog/HadrichBPPM21} and describe mass reduction of a Gaussian element as
\begin{equation} \label{eq:mainmassloss}
    \Delta m = - k(T) c \bar{A}\,,
\end{equation}
where $m$ is the remaining mass, $k(T)$ is the temperature-dependent reaction rate [$\mathrm{s^{-1}}$], and $c$ denotes the charring mass per unit surface area $[\mathrm{kg}\,\mathrm{m}^{-2}]$\,.
We adopt the specification of the charring coefficient $c$ from the same work.

We treat this coefficient as a spatially constant parameter, as char-layer dynamics are not explicitly modeled. Let $\bar{A}$ be the effective pyrolysing surface area [$\mathrm{m^2}$]. Following Pirk et al. \shortcite{DBLP:journals/tog/PirkJHMP17}, we define the reaction rate $k(T_g)$ using onset and full-activation temperatures $T_0 = 150^\circ\mathrm{C}$ and $T_1 = 450^\circ\mathrm{C}$ as
\begin{equation}
   k(T_g) = 
   \begin{cases} 
        0 & \text{if } T_g < T_0, \\
        S\left(\frac{T_g - T_0}{T_1 - T_0}\right) & \text{if } T_0 \leq T_g \leq T_1,\\
        1 & \text{if }  T_g > T_1,
    \end{cases}
\end{equation}
where $S(x) = 3x^2 - 2x^3$ is a sigmoid interpolation function.

\section{Evaluation\label{sec:results}}

\begin{figure*}[!ht]
    \centering
    \subfloat[\label{fig:spread_slopes_d20} $\theta = -20 \degree$]{\includegraphics[width=0.19\textwidth]{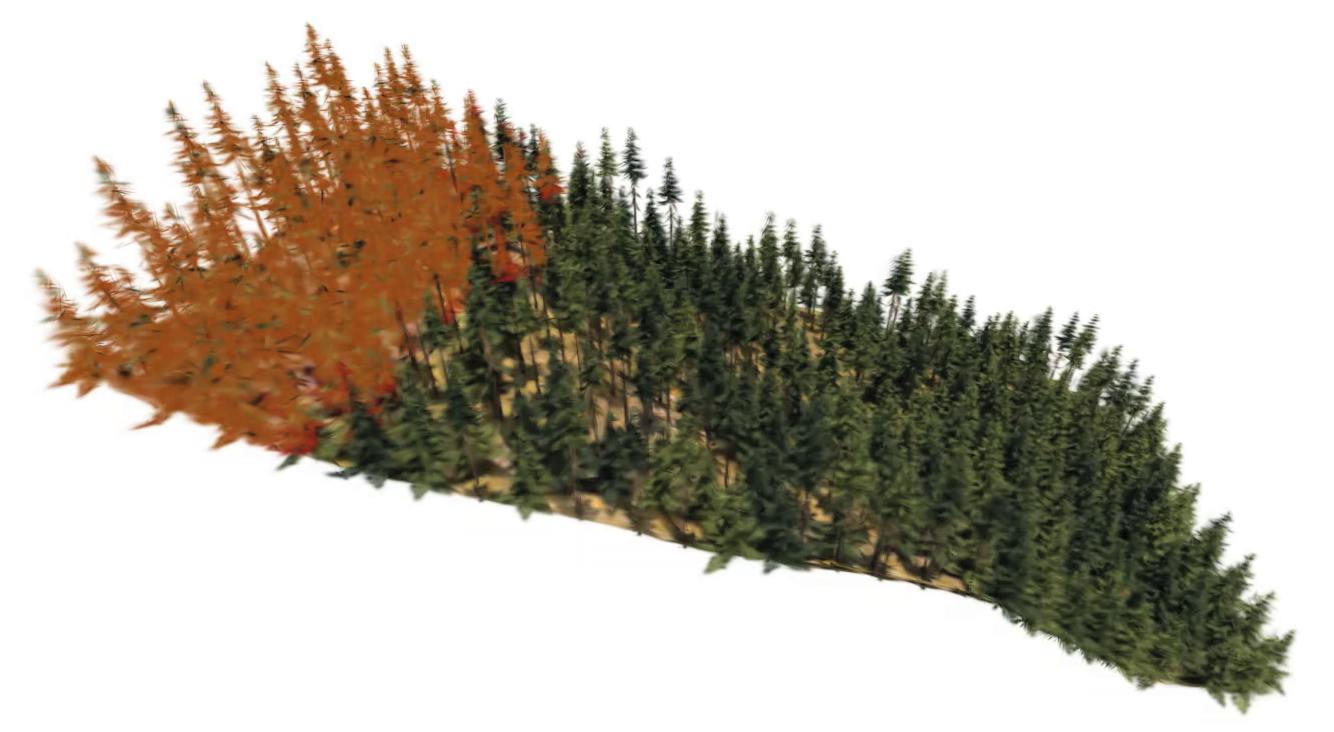}} \hfill
    \subfloat[\label{fig:spread_slopes_d10} $\theta = -10 \degree$]{\includegraphics[width=0.19\textwidth]{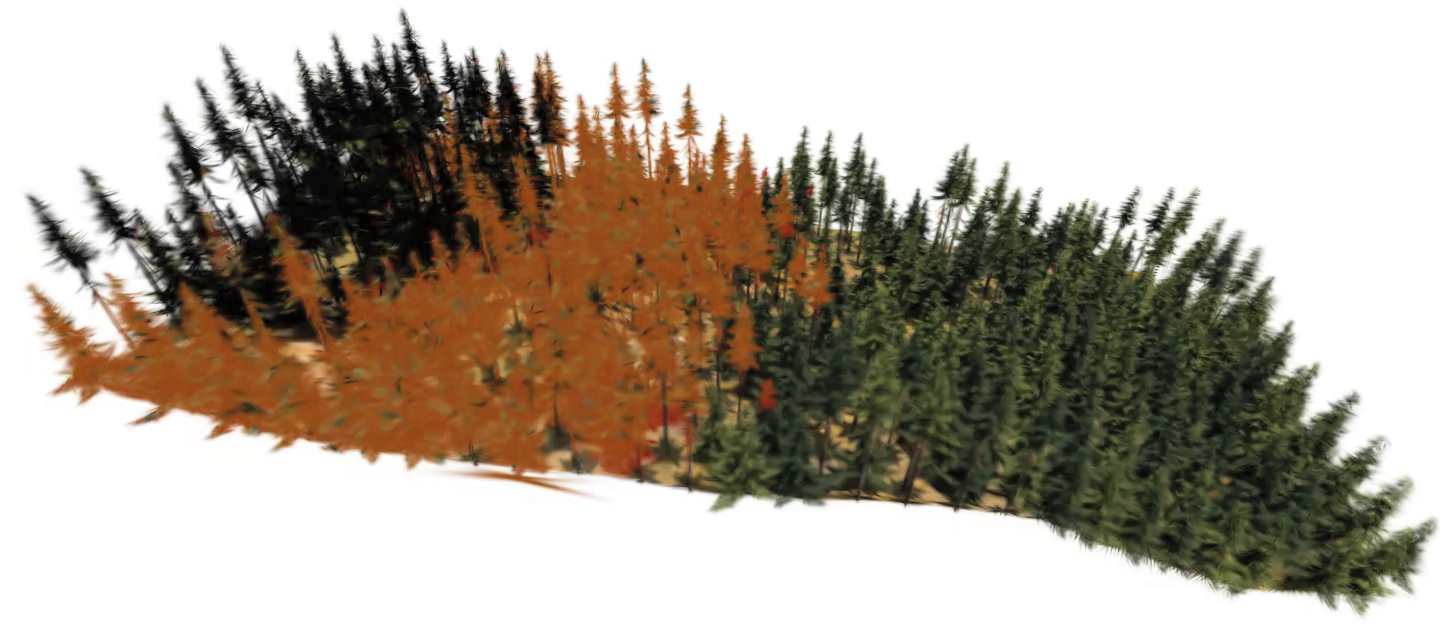}} \hfill
    \subfloat[\label{fig:spread_slopes_0} $\theta = 0 \degree$]{\includegraphics[width=0.19\textwidth]{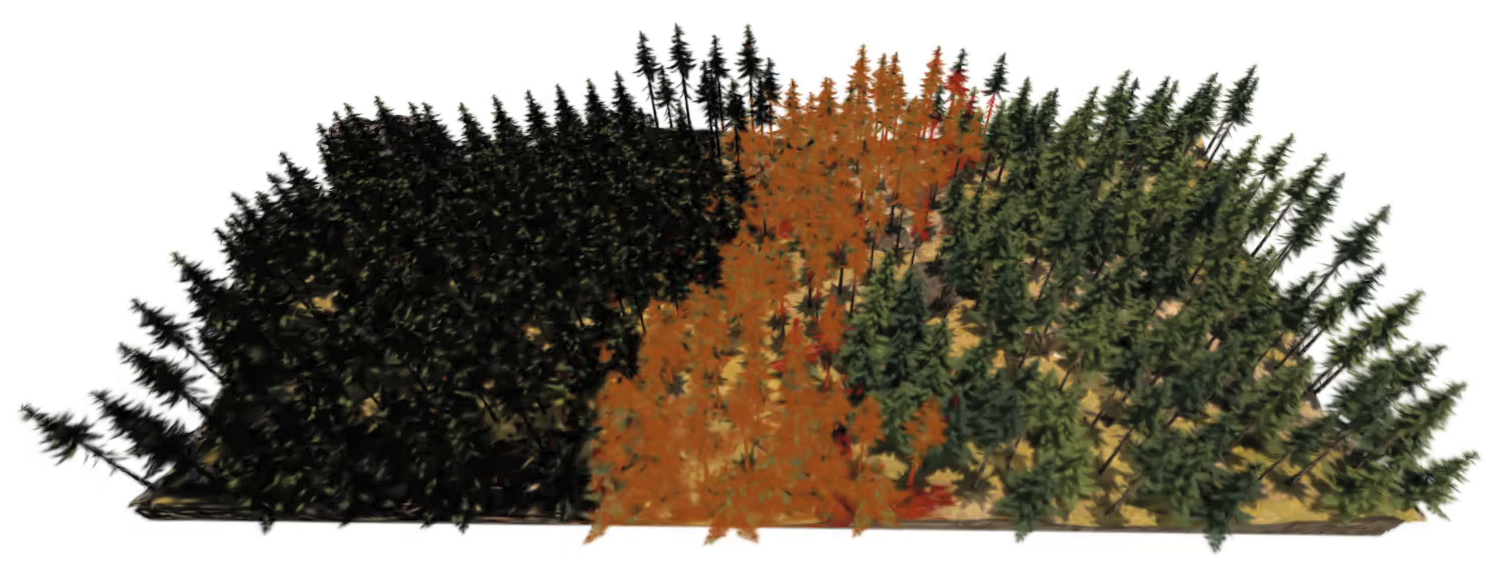}} \hfill
    \subfloat[\label{fig:spread_slopes_u10} $\theta = 10 \degree$]{\includegraphics[width=0.19\textwidth]{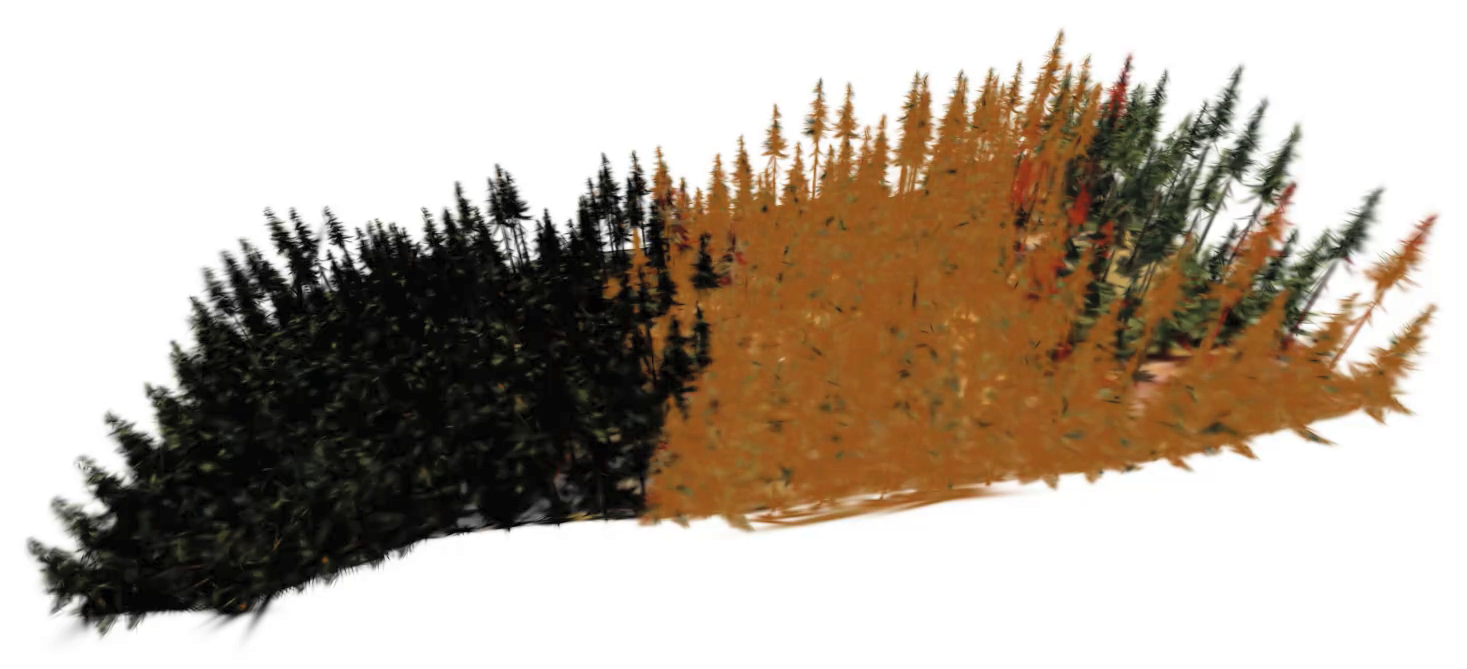}} \hfill
    \subfloat[\label{fig:spread_slopes_u20} $\theta = 20 \degree$]{\includegraphics[width=0.19\textwidth]{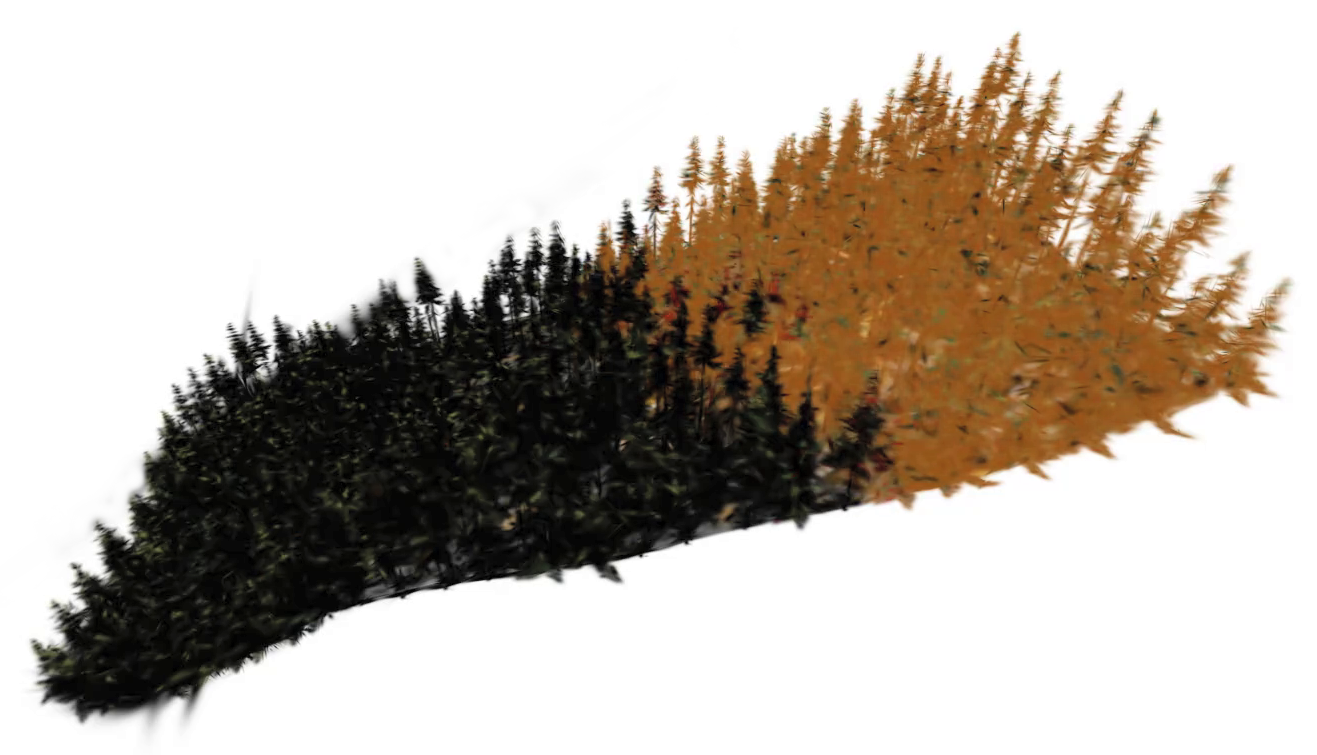}}
    \caption{Fire spread across five terrains with different slope angles$\theta$, ranging from $-20\,\degree$ (downhill) to $+20\,\degree$ (uphill). Each scene spans $50 \times 100$\,m. In all cases, ignition occurred at the same $(x,y)$ 
    % location at the lower end of the slope, 
    and the wind magnitude was zero, so that any anisotropy in spread can be attributed purely to the slope.}
    \label{fig:spread_slopes}
\end{figure*}

We implemented our method on top of Nerfstudio~\cite{tancik2023nerfstudio}, leveraging its built-in Feature Splatting, as well as CUDA-accelerated PyTorch code for efficient simulation.
For our experiments, we use a machine equipped with an NVIDIA RTX 3090.

We evaluate our method on both synthetic and real scenes.
For the creation of synthetic scenes, we placed realistic assets on a vertically displaced subdivided quad to represent terrain, which we then rendered from aerial viewpoints simulating drone capture.
Furthermore, we use real-world data captured by the Open Forest Observatory \cite{openforestobservatory-2026}.
Both settings are processed in the same manner using COLMAP for structure-from-motion~\cite{Schonberger_2016_CVPR} and trained as 3DGS scenes with semantic feature embeddings~\cite{DBLP:journals/tog/GaussianSplatting,DBLP:conf/eccv/QiuYZW24}.

\begin{figure}
    \subfloat[\label{fig:spread_density_25}$25\%$ density]{\includegraphics[width=0.32\columnwidth]{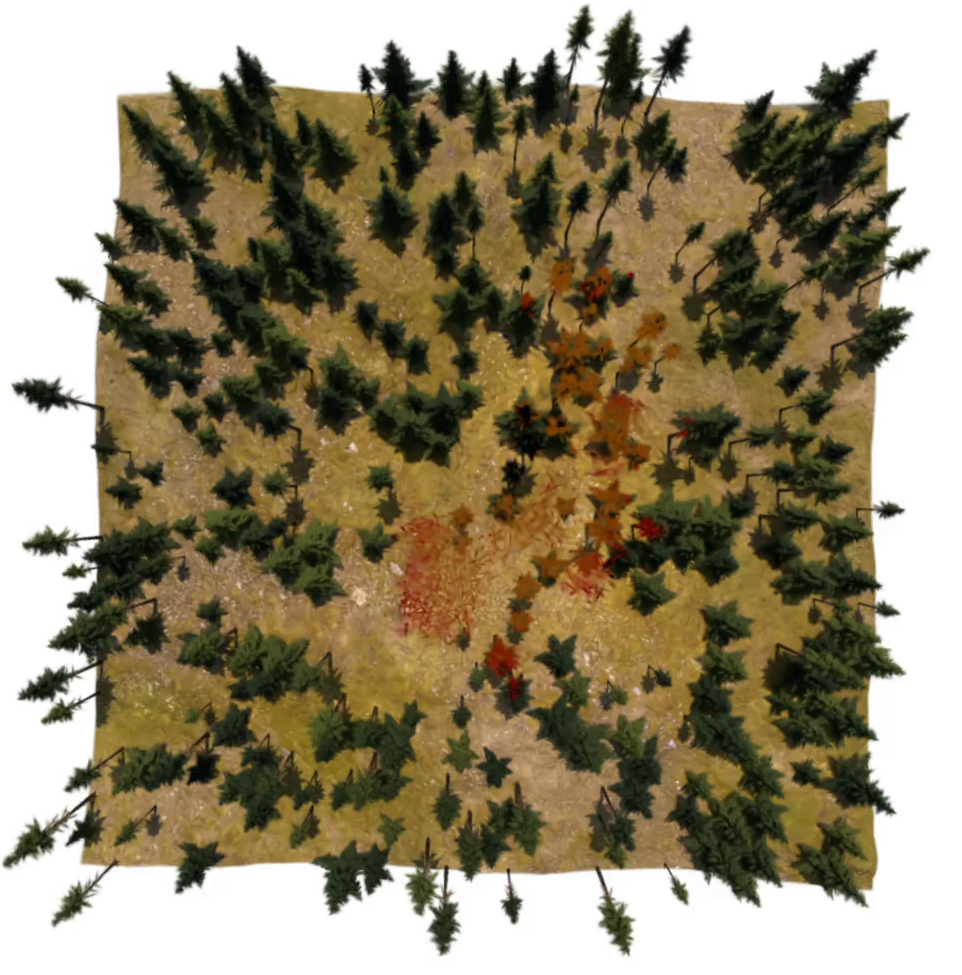}}
    \subfloat[\label{fig:spread_density_45}$45\%$ density]{\includegraphics[width=0.32\columnwidth]{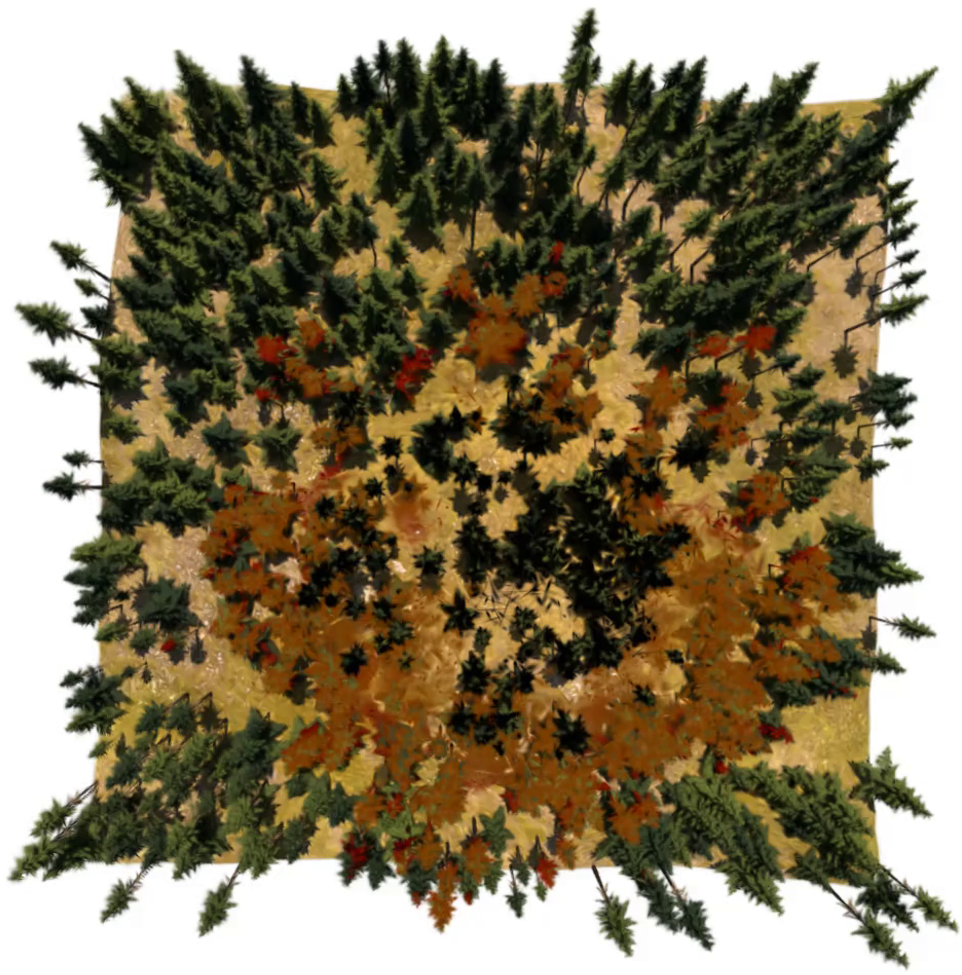}}
    \subfloat[\label{fig:spread_density_90}$90\%$ density]{\includegraphics[width=0.32\columnwidth]{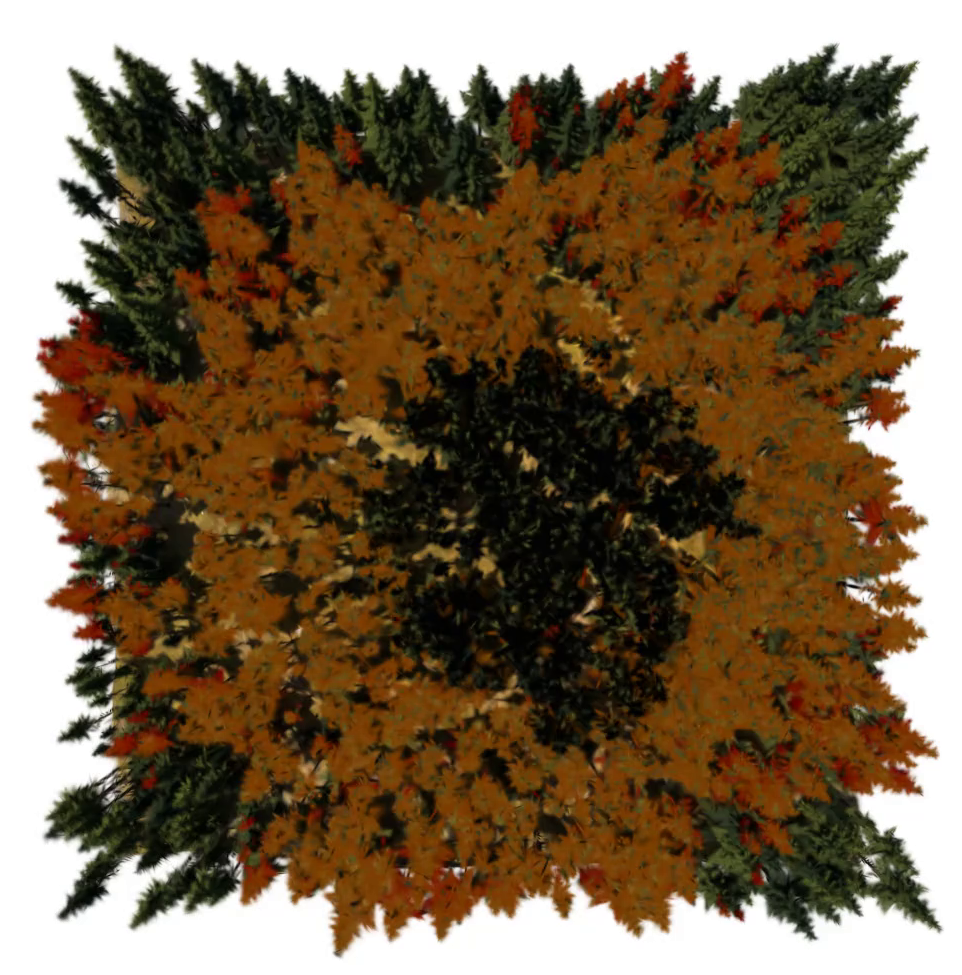}}
    \caption{\label{fig:spread_density}
    Fire spread at the same time step shown for three different scene densities (25\%, 45\%, and 90\% vegetation cover). At higher densities, the fire front spreads faster and more continuously through the available fuel.
    }
\end{figure}

\begin{figure}
    \includegraphics[width=\columnwidth]{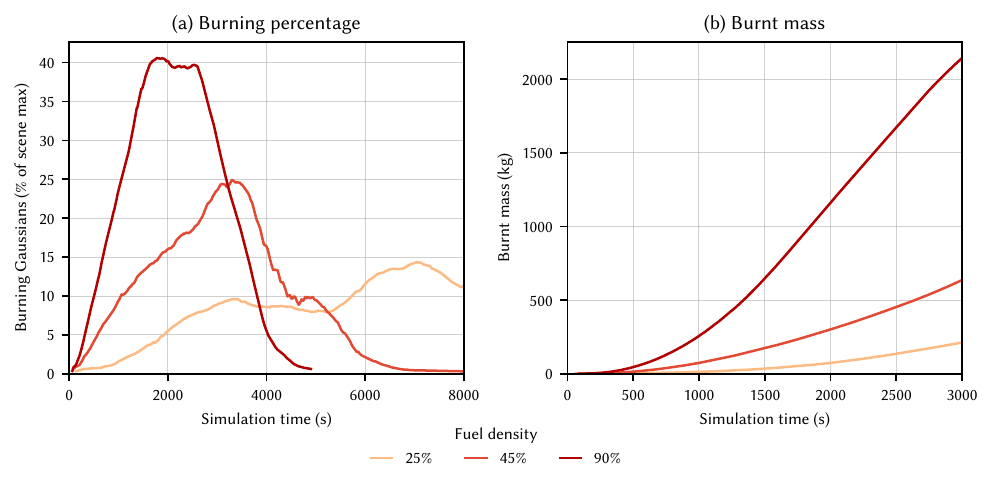}
    \caption{\label{fig:density_burning_stats} Quantitative analysis of fire dynamics for varying vegetation densities. (a) Percentage of currently burning Gaussians over time, showing that denser forests ignite faster but also exhaust their fuel earlier. (b) Cumulative burnt mass over time for the same scenes, scaling super-linearly with vegetation density while fuel is still available.}
\end{figure}

\paragraph{\textbf{Fire Spread at Varying Vegetation Density}}
We analyze fire spread across various synthetic scenes, for which we created scenes with varying vegetation densities (25\%, 45\%, and 90\%), where we distributed trees via Poisson-disk sampling and varied their height using fractal noise.
Fig.~\ref{fig:spread_density} shows exemplary scenes for different densities and demonstrates how fire spreads after the same elapsed time.
We observe that higher fuel density leads to faster, more continuous fire propagation, where the fire in a 90\%-density forest progresses faster through the scene. This behavior aligns with real-world wildfire dynamics, where fuel availability strongly governs the spread rate.
Fig.~\ref{fig:density_burning_stats} (a) shows the percentage of burning Gaussians over time, demonstrating that denser forests burn faster and exhaust fuel more rapidly. Sparser scenes burn more gradually, but less consistently. Fig.~\ref{fig:density_burning_stats} (b) shows that the cumulative burned mass scales super-linearly with density, correctly indicating that dense regions of forest pose an even greater risk for rapid fire spreading.

\begin{figure}%[p]%[!ht]
    \subfloat[\label{fig:graph_slope_spread_speed} Propagation speed per angle]{\includegraphics[width=0.49\columnwidth]{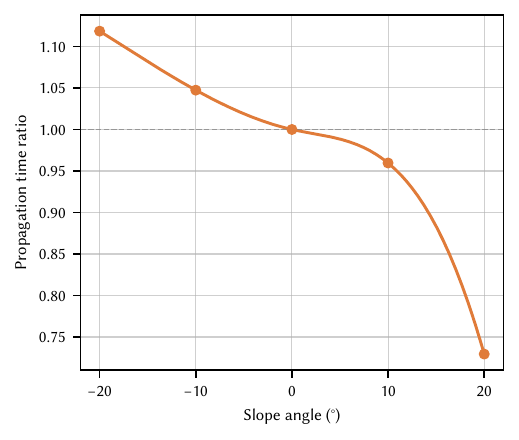}}
    \subfloat[\label{fig:graph_simulation_consistency} Simulation consistency]{\includegraphics[width=0.49\columnwidth]{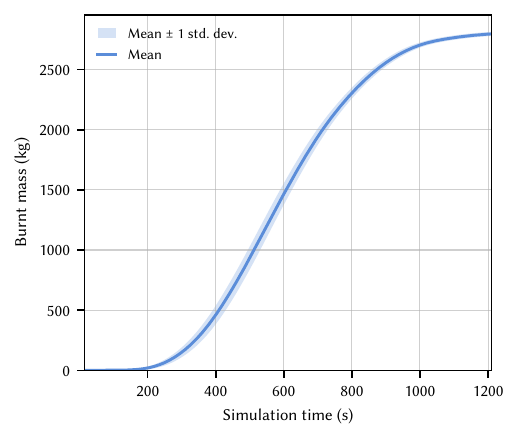}}
    \caption{(a) and (b) summarise the quantitative behaviour of our simulation. (a) Fire propagation times for varying slope angles, relative to no incline. We see that upward inclines correspond to drastically faster fire propagation, with a clearly non-linear acceleration at steeper angles, while downhill spread responds more linearly. (b) Simulation consistency across 5 separate burns on the same scene. The shaded band indicates the standard deviation around the mean cumulative burnt mass, illustrating that the stochastic particle dynamics yield reproducible aggregate behaviour.}
\end{figure}

\paragraph{Terrain Incline and Slope}
To verify physical accuracy, we evaluate the relationship between terrain incline and propagation speed, replicating experiments from established models \cite{YOU2022109, DBLP:journals/tog/HadrichBPPM21}. It is a well-documented phenomenon that wildland fires accelerate uphill and decelerate downhill. As illustrated in Fig.~\ref{fig:spread_slopes}, our model accurately captures the acceleration of uphill fires and deceleration of downhill spread under zero-wind conditions. Quantitatively, Fig.~\ref{fig:graph_slope_spread_speed} shows a non-linear relationship between slope steepness and uphill spread rate, aligning with Rothermel’s Rate of Spread (ROS) model \cite{rothermel1972mathematical}. Our simulation captures the empirical non-linearity noted by Geng et al. \shortcite{10.1071/WF23118}, where ROS increases disproportionately at steeper angles (e.g., $10^\circ$ to $30^\circ$). This suggests that our framework's modular interaction logic effectively approximates complex topographical fire-atmosphere coupling.

\begin{figure}%[p]%[!ht]
    % --- Row 1: Wind Speed 5 m/s ---
    \subfloat{\includegraphics[width=0.32\columnwidth]{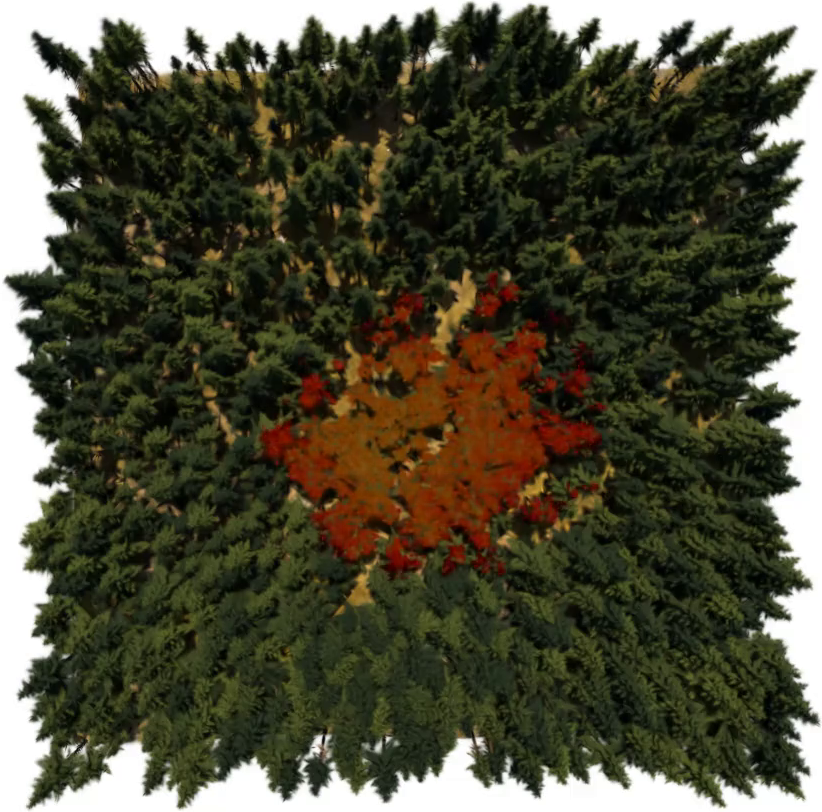}} \hfill
    \subfloat{\includegraphics[width=0.32\columnwidth]{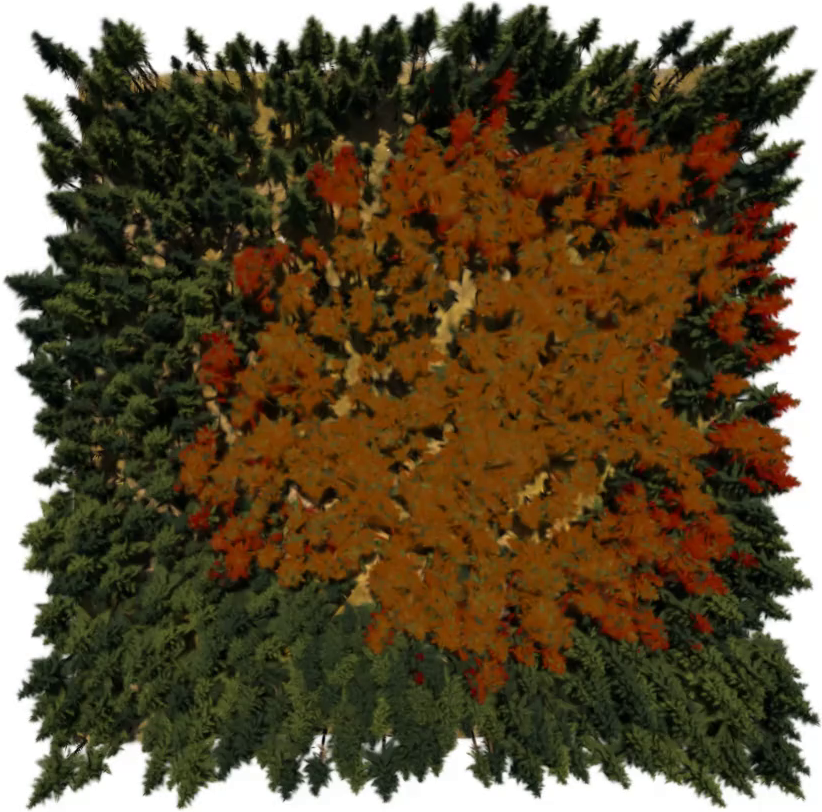}} \hfill
    \subfloat{\includegraphics[width=0.32\columnwidth]{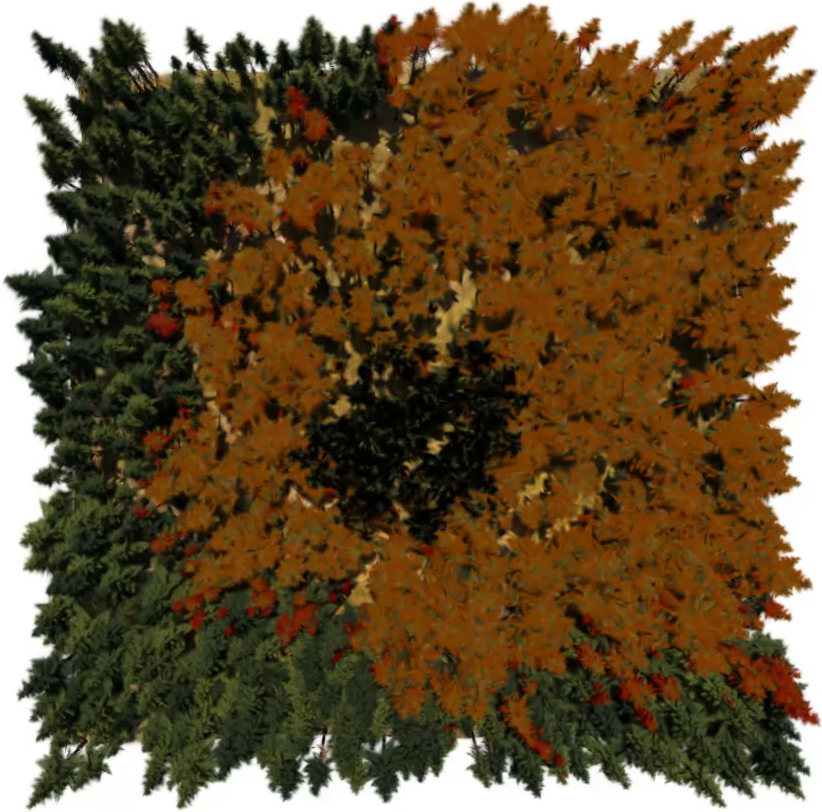}} \\[-2.5ex] 

    % --- Row 2: Wind Speed 15 m/s ---
    \subfloat{\includegraphics[width=0.32\columnwidth]{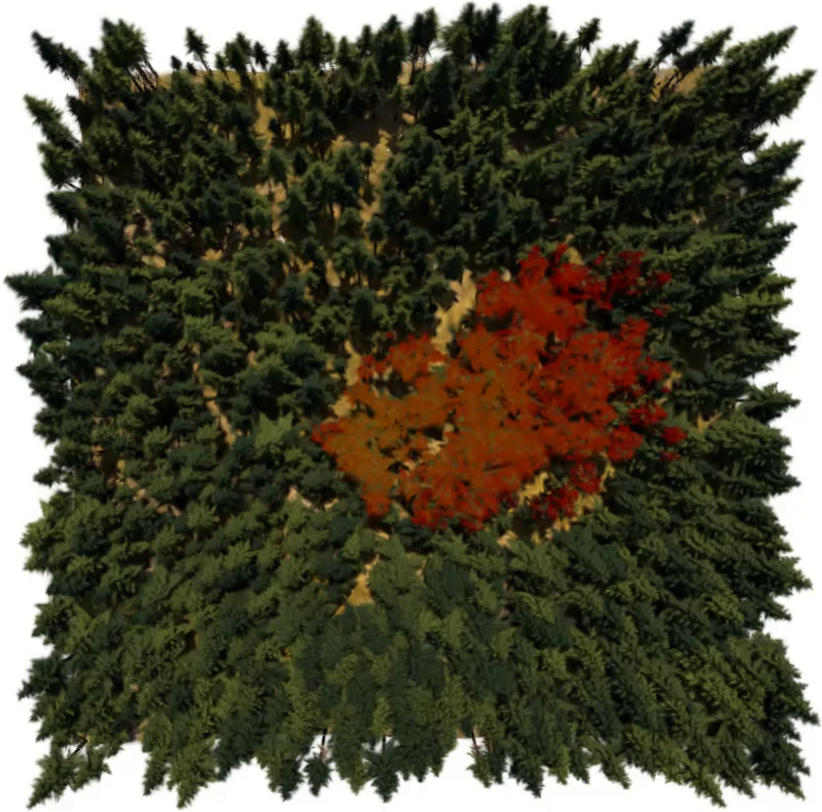}} \hfill
    \subfloat{\includegraphics[width=0.32\columnwidth]{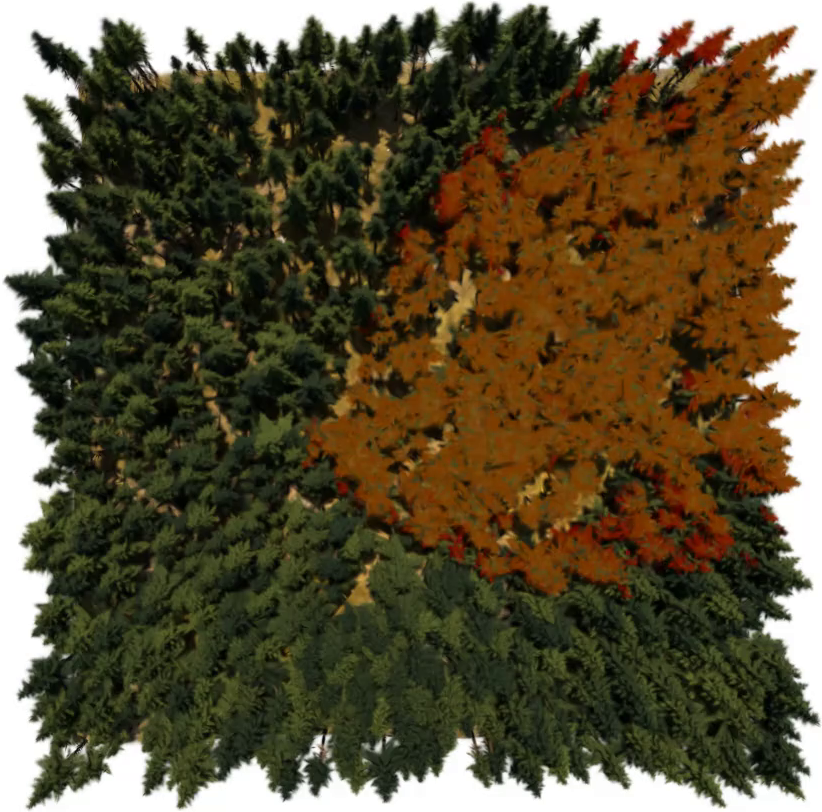}} \hfill
    \subfloat{\includegraphics[width=0.32\columnwidth]{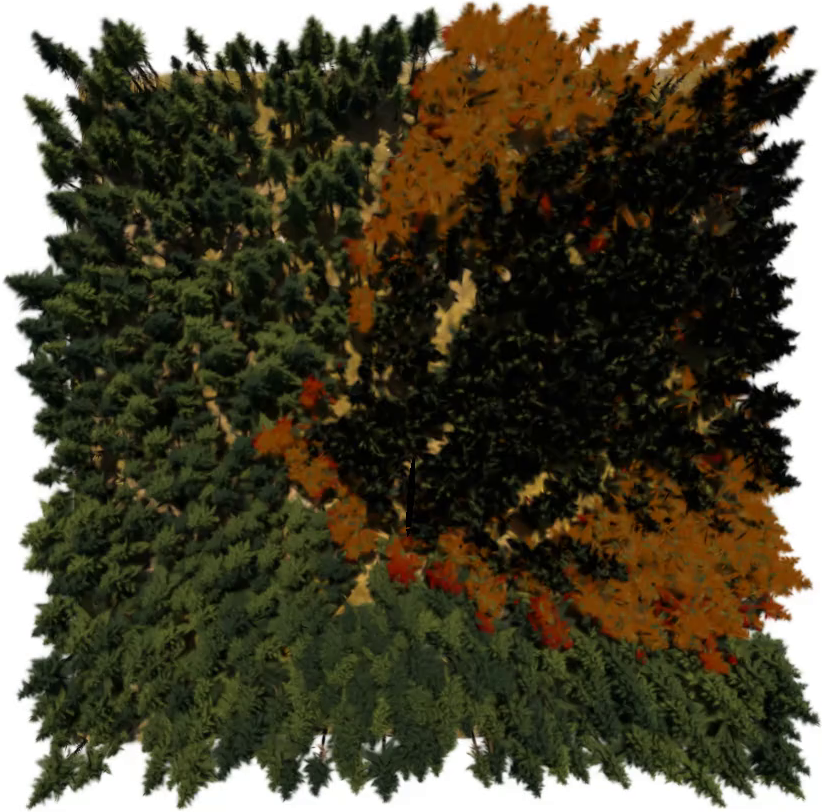}}
    
    \caption{Fire spread comparison under varying wind speeds on a synthetic forest scene ($60 \times 60$\,m terrain, $90\%$ vegetation density, identical ignition point and wind direction). Top row: $5\,\mathrm{m/s}$ wind; bottom row: $15\,\mathrm{m/s}$ wind. The three columns correspond to snapshots taken at 5\,minutes (left), 15\,minutes (middle), and 30\,minutes (right) of simulated time. Higher wind speeds result in significant advection, elongating the fire front and accelerating fuel consumption visibly across the same time intervals.}
    \label{fig:wind_speed_effect}
\end{figure}

\paragraph{\textbf{Fire Spread at Varying Wind Speeds}}
We test the wind model using a 90\%-density forest scene to provide a uniform fuel bed and minimize the impact of tree positioning. As shown in Fig.~\ref{fig:wind_speed_effect}, varying the wind speed while keeping the ignition point and direction constant results in a distinct anisotropic spread. Specifically, higher wind velocities produce more elongated fire fronts and faster burnout rates, validating that our implementation correctly scales the rate of spread with wind strength.

\paragraph{\textbf{Consistency}}
Our simulation is not deterministic due to the randomized particle movement. Therefore, we assess the consistency of our forest fire model over five runs. Fig.~\ref{fig:graph_simulation_consistency} shows that the cumulative burnt mass over time is nearly identical for all runs, with only a minor deviation that decreases towards the end.

\paragraph{\textbf{Simulating Fuel Discontinuities and Fire Breaks}}
We evaluate the ability of our method to handle fuel discontinuities by simulating both natural barriers (e.g., riverbeds) and standardized firefighting interventions. We define firebreaks as strips cleared to mineral soil (width $W_{\text{fire}} = 3H_{\text{fuel}}$) and fuelbreaks as wider zones of reduced vegetation density. To account for slope-driven flame extension and wind-driven spotting, the fuelbreak width is dynamically defined as $W_{\text{fuel}} = \max(W_{\text{slope}}, W_{\text{wind}}, 3H_{\text{fuel}})$\,.

\begin{figure}%[p]%[!ht]
    \centering
    % --- Row 1: Wind Speed 0 m/s ---
    \subfloat{\includegraphics[width=0.32\columnwidth]{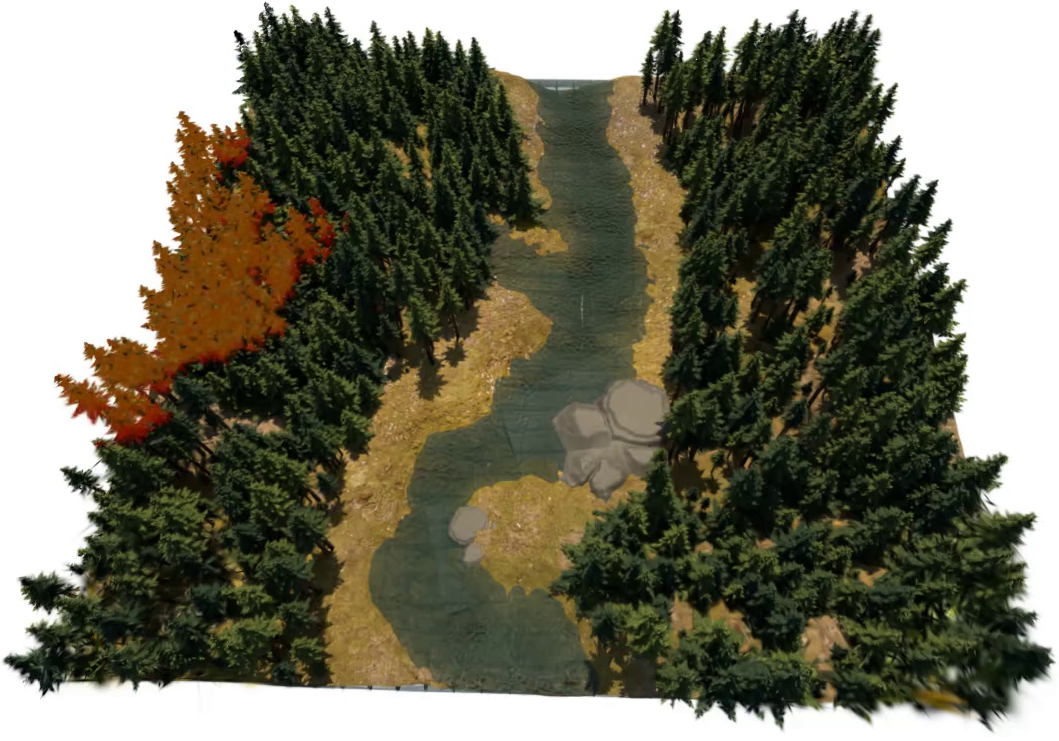}} \hfill
    \subfloat{\includegraphics[width=0.32\columnwidth]{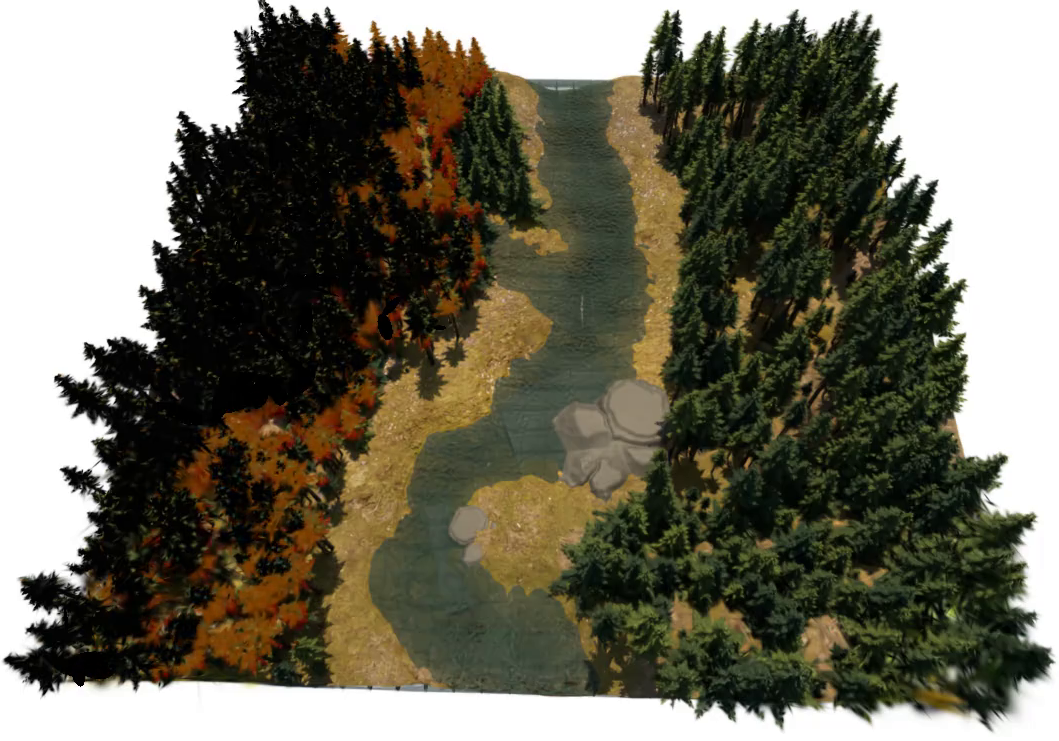}} \hfill
    \subfloat{\includegraphics[width=0.32\columnwidth]{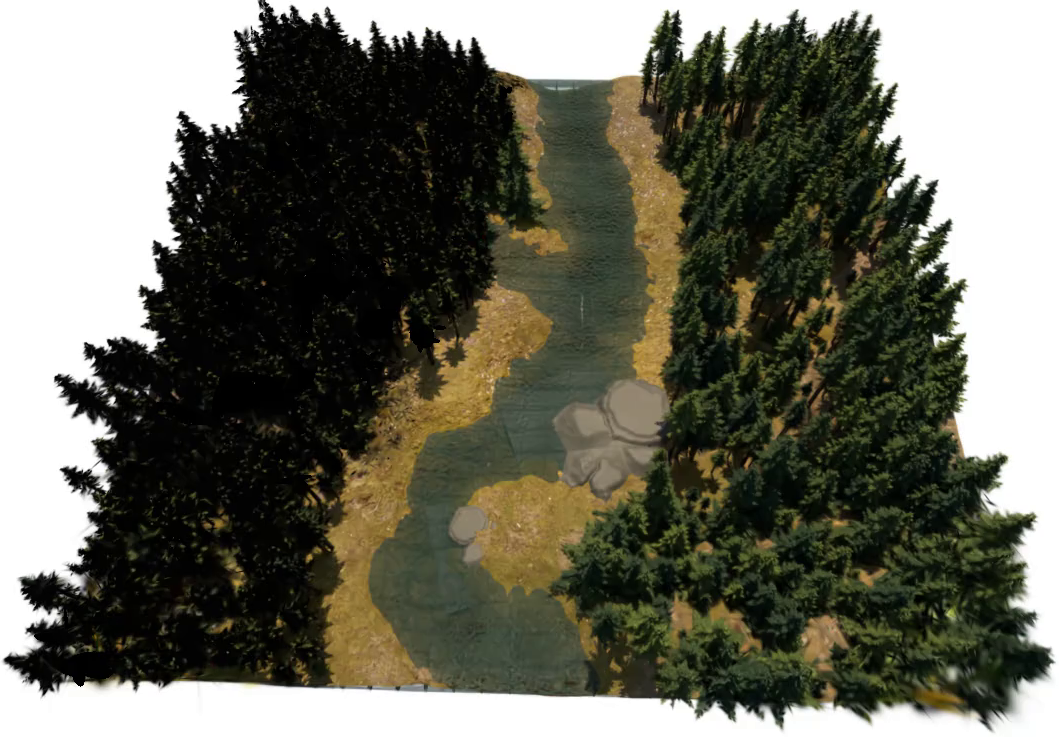}} \\[-2.5ex] 

    % --- Row 2: Wind Speed 21 m/s ---
    \subfloat{\includegraphics[width=0.32\columnwidth]{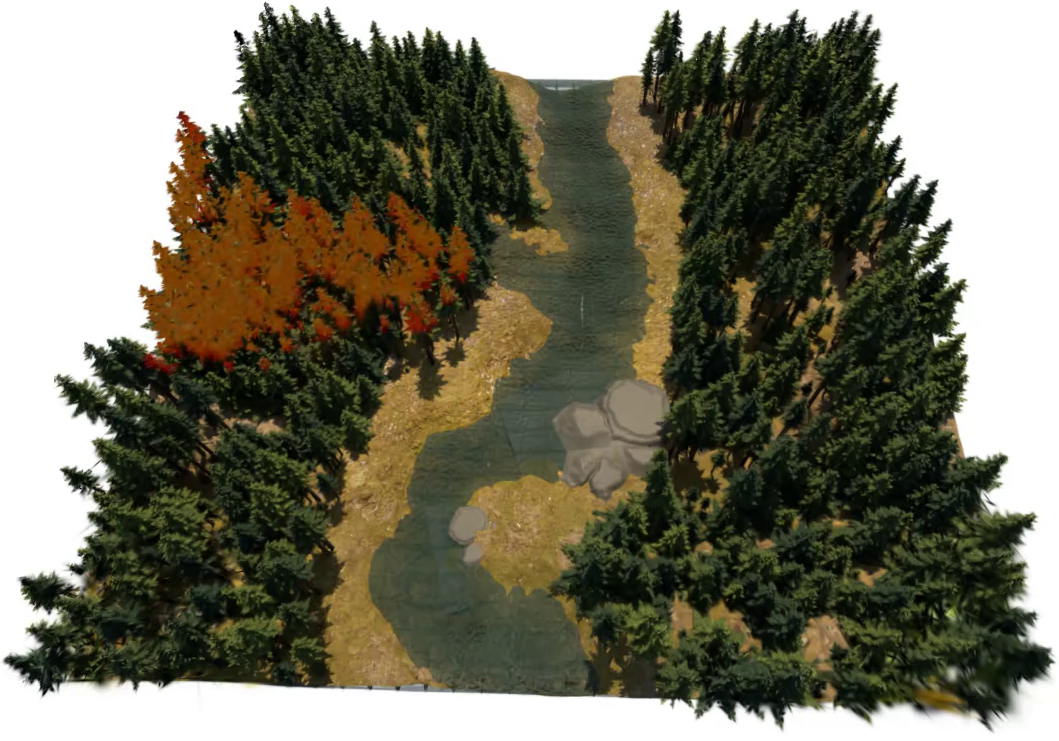}} \hfill
    \subfloat{\includegraphics[width=0.32\columnwidth]{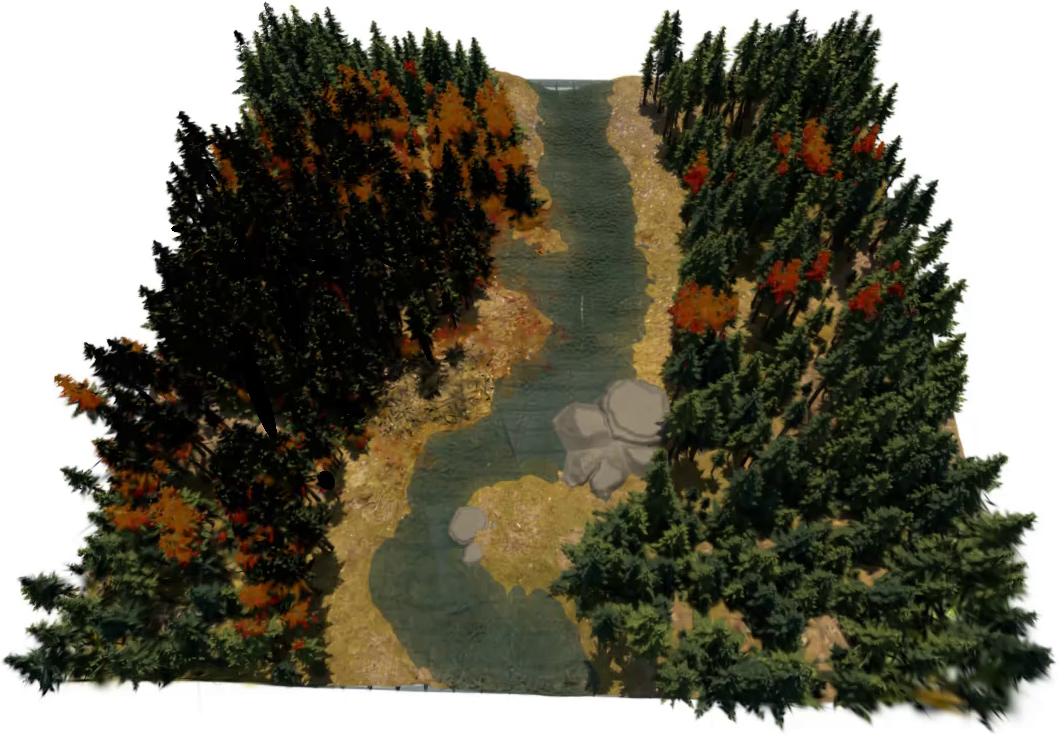}} \hfill
    \subfloat{\includegraphics[width=0.32\columnwidth]{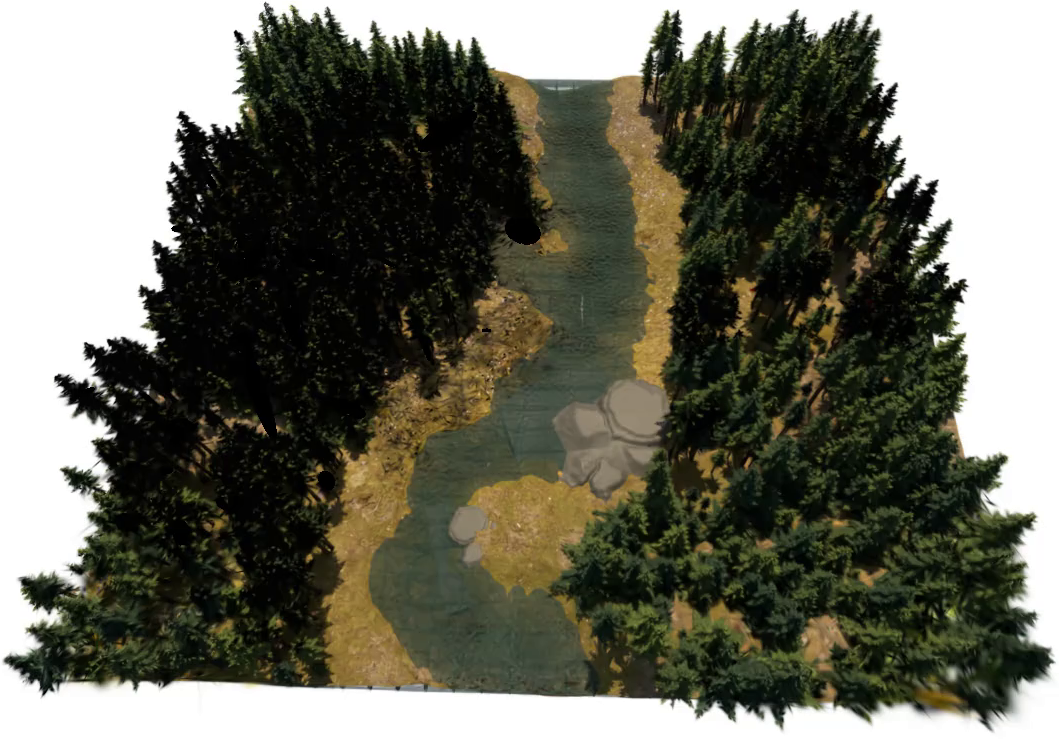}}
    
    \caption{Example of a river acting as a natural firebreak stopping the fire spread in a synthetic scene of $170 \times 170$\,m extent. Three time steps are shown per row (early, intermediate, and late). Under zero-wind conditions (top row), the fire is contained by the river. In the case of strong wind blowing to the right, shown here at $21\,\mathrm{m/s}$ (bottom row), the fire leaps across the gap via wind-driven spotting and elevated heat flux.}
    \label{fig:natural_fire_break}
\end{figure}

\begin{figure}%[p]%[!ht]
    \includegraphics[width=\columnwidth]{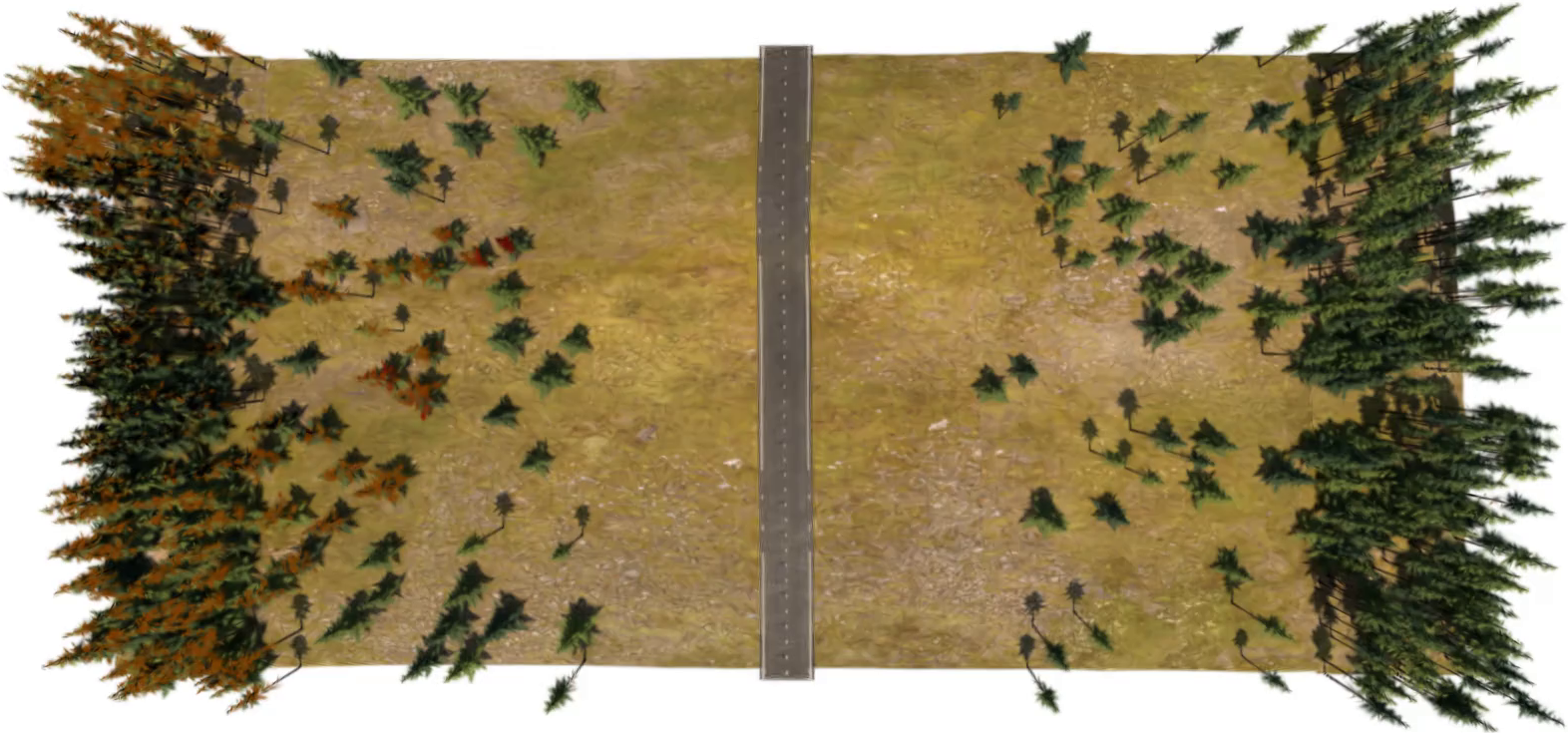}
    \caption{\label{fig:fire_break_spread_no_wind} Firebreak simulation with no wind on a synthetic forest scene with a centrally placed road. Despite the dense vegetation on both sides, the fire fails to cross from the left side to the right side, demonstrating that the firebreak is wide enough to halt propagation under calm conditions. The road is explicitly classified as nonflammable, and any subterrain Gaussians are excluded from combustion.}
\end{figure}

As shown in Figs.~\ref{fig:natural_fire_break} and \ref{fig:fire_break_spread_no_wind}, these discontinuities effectively halt propagation under zero-wind conditions. However, the simulation captures the critical threshold where extreme wind enables the fire to leap across gaps via increased heat flux and spotting, a key requirement for high-fidelity wildfire modeling.
To ensure these results reflect fuel dynamics rather than classification noise, all Gaussians identified as being below the terrain height map (Sec.~\ref{subsec:scene_processing}) are explicitly treated as non-flammable.

\begin{figure}%[p]%[!ht]
    \includegraphics[width=\columnwidth]{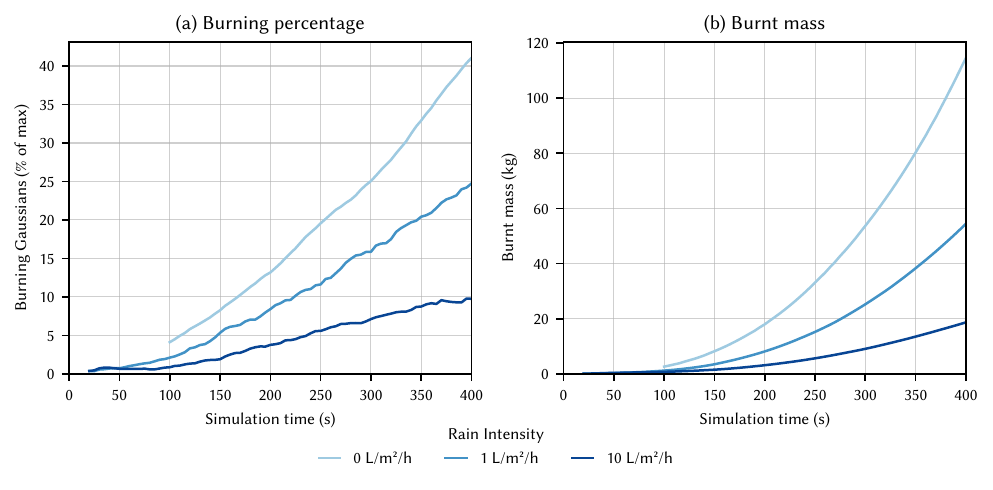}
    \caption{\label{fig:rain_burning_stats} Effect of rain-driven cooling on the fire dynamics. (a) Percentage of currently burning Gaussians and (b) cumulative burned mass over time, for three different rain intensities ranging from no rain to heavy rain. Even light rain noticeably slows down the combustion process, while heavy rain produces erratic, intermittent spread.}
\end{figure}

\paragraph{\textbf{Rain Particles}}
Fig.~\ref{fig:rain_burning_stats} shows that an increase in rain has a clear effect on burnt mass over time. Heavy rain ($10\,\mathrm{L}\,\mathrm{m}^{-2}\,\mathrm{h}^{-1}$) results in a much less consistent fire spread due to the constant cooling of burnable Gaussians. We notice that even light rain ($1\,\mathrm{L}\,\mathrm{m}^{-2}\,\mathrm{h}^{-1}$) has a significant effect on the combustion process.

\paragraph{\textbf{Real-World Scenes}}
We evaluated the framework on in-the-wild data from the Open Forest Observatory dataset \cite{openforestobservatory-2026}. The LiDAR and camera extrinsics were normalized to the local coordinate frame of the simulation to ensure consistent scale and orientation. As shown in Fig.~\ref{fig:teaser}, the model successfully simulates fire spread through the reconstructed vegetation.

\paragraph{\textbf{Discussion}}
The quality of our simulation is inherently coupled to the quality of the input and output of the 3DGS reconstruction. Standard 3DGS artifacts, such as view-dependent density variations and irregular primitive distributions at scene boundaries, can introduce variance into local combustion rates. Additionally, an incorrect global scale (i.e., $1$\,m in the real world not corresponding to $1$ unit in the simulation) can result in incorrect simulations.
Furthermore, the scene remains geometrically static after combustion since the Gaussians are only marked as burnt. The scene geometry does not yet undergo physical deformation or structural collapse during the simulation, which we leave for future work.
Current semantic classification also lacks the granularity to distinguish between live and dead fuel (e.g., dry vs. green branches), which is a critical factor in ignition thresholds. Additionally, while we envision a ``digital twin'' application, live-updating the simulation during active fires remains a significant challenge, as smoke and thermal turbulence interfere with standard visual reconstruction. Finally, our evaluation focuses primarily on homogeneous boreal forests, but we foresee that further tuning will make our method applicable to heterogeneous environments with more complex tree compositions.
\section{Conclusion\label{sec:conclusion}}

We have presented WildFireGS, a modular wildfire simulation framework that leverages the semantic and spatial properties of 3D Gaussian Splatting to enable physics-based simulation of wildfire behavior in captured forest environments. Our evaluation demonstrates that our model reproduces characteristic wildfire dynamics across both synthetic and real-world forest reconstructions. Controlled experiments verify that the simulation responds realistically to environmental variables and can be extended to integrate other processes such as rain-driven cooling mechanisms.

Future research could explore more sophisticated structural modeling of vegetation, moving beyond per-Gaussian mass calculations toward aggregate physical properties for entire tree structures. 
Moreover, extending the system to support multi-layer dynamics would more accurately capture the complex vertical fuel structures and reignition phenomena characteristic of real-world forest fires.
\clearpage

\clearpage

\bibliographystyle{unsrt}  
\bibliography{references}  %%% Remove comment to use the external .bib 

\end{document}